\documentclass[floatfix,10pt,aps,prb,twocolumn,a4paper]{revtex4}
\usepackage{amsmath}   % contains the advanced math extensions for LaTeX
\usepackage{mathtools} % an extension package to amsmath
\usepackage{amssymb}   % adds new symbols in to be used in math mode
\usepackage{mathrsfs}  % other mathematical symbols
\usepackage{textcomp}  % provides extra symbols
\usepackage{accents}   % provides some miscellaneous tools for mathematical accents
\usepackage{bm}        % provides bold font for math symbols
\usepackage{soul}
\usepackage{relsize}
\usepackage{graphicx}  % to manage external pictures
\usepackage{epstopdf}  % adds support of handling eps images to package graphics or graphicx
\usepackage{subfigure} % provides support for the inclusion of small, ‘sub’, figures and tables
\usepackage{makecell}
\usepackage{hhline}
\usepackage{cellspace}
\usepackage{array}     % extends the possibility of LaTeX to create very complicated and customized tables
\usepackage{tabularx}  % modifies the widths of certain columns, rather than the inter column space
\usepackage{multirow}  % provides a construction for table cells that span more than one row of the table
\usepackage{booktabs}  % publication quality tables
\usepackage{dcolumn}   % defines a system for defining columns of entries in an array or tabular which are to be aligned on a 'decimal point'
\usepackage{gensymb}   % for \degree and \celcius
\usepackage{setspace}
\usepackage{scrextend}
\usepackage{multirow}

\usepackage[usenames,dvipsnames]{color} % to use color text in LaTeX2e
\usepackage{url}       % defines the \url{...} command
\usepackage{hyperref}  % extends the functionality of all the LaTeX cross-referencing commands
\hypersetup{colorlinks,citecolor=Violet,linkcolor=Red,urlcolor=Blue}

\graphicspath{{./}}
\begin{document}
\setlength{\heavyrulewidth}{0.08em}
\setlength{\lightrulewidth}{0.05em}
\setlength{\cmidrulewidth}{0.03em}
\setlength{\belowrulesep}{0.65ex}
\setlength{\belowbottomsep}{0.00pt}
\setlength{\aboverulesep}{0.40ex}
\setlength{\abovetopsep}{0.00pt}
\setlength{\cmidrulesep}{\doublerulesep}
\setlength{\cmidrulekern}{0.50em}
\setlength{\defaultaddspace}{0.50em}
\setlength{\tabcolsep}{4pt}
\title{Orbital and magnetic ordering in single-layer FePS$_{3}$: A DFT$+U$ study}
\author{Mohammad Amirabbasi}\email {mo.amirabbasi@gmail.com}
\author{Peter Kratzer}\email {Peter.Kratzer@uni-due.de}
\affiliation{Fakult\"at f\"ur Physik and CENIDE, Universit\"at Duisburg-Essen, Lotharstra{\ss}e 1, 47057 Duisburg, Germany}
\date{\today}
%---------- ABSTRACT --------------
\begin{abstract}
Among the numerous 2D system that can be prepared via exfoliation,
iron phosphorus  trisulfide (FePS$_{3}$) attracts a lot of attention 
recently due to its broad-range photoresponse, its unusual Ising-type magnetic order and possible applications in spintronic nano-devices. Despite various experimental and theoretical-computational reports, there are still uncertainties in identifying its magnetic ground state. In this paper, we investigate the structural and magnetic properties of single-layer FePS$_{3}$ by using Density Functional Theory.  Our findings show that 
orbital ordering leads to a variation  in distance between pairs of iron atoms by 0.14~\AA. These lattice distortions, albeit small,  
trigger different (ferromagnetic and antiferromagnetic) 
exchange couplings so that the ground state consists of ferromagnetically aligned zigzag chains along the long Fe -- Fe bonds which couple antiferromagnetically along the shorter Fe -- Fe bonds. 
Within the DFT$+U$ framework, we parameterize a spin Hamiltonian including Heisenberg, single-ion anisotropy, Dzyaloshinskii-Moriya and biquadratic interactions. 
Using $U=2.22$~eV gives a consistent description of both the electronic band gap and the Neel temperature in 2D FePS$_3$.
\end{abstract}
\maketitle

%%%%%##################################################################################################################
%%%%%##################################################################################################################
%%%%%##################################################################################################################
\section{introduction}
\label{sec:introduction}
The study of two-dimensional (2D) magnetic ground states has gained special interest after the discovery of stable long-range ferromagnetic (FM) and antiferromagnetic (AFM) order in monolayer CrI$_{3}$ and FePS$_{3}$, respectively\cite{Huang2017, Gong2017}.  The ideal candidates for 2D magnets are layered van der Waals 
materials such as  the transition-metal dichalcogenides\cite{Chhowalla2013},  chromium trihalides\cite{Huang2017} and transition-metal phosphorous trichalcogenides~\cite{Zhang2015}. 
These materials hold significant promise for technological applications, especially in the field of spintronics and nanomagnetism~\cite{Ahn2020}. Therefore, a deep understanding of the magnetic exchange mechanisms in 2D system is essential. In this way, it is possible to find out which exchange coupling controls the magnetic properties of the ground state and how the system selects a special order when decreasing the temperature. Here,  we investigate the magnetic ordering in FePS$_{3}$, a 2D Ising antiferromagnet with a Neel temperature of 116 -- 120 K in bulk~\cite{lee2016ising, wang2016raman}. 
In advantage over other 2D materials, Ramos {\it et al.}~\cite{Ramos2021} claimed the 
optical response of FePS$_3$ over a broad range of the electromagnetic spectrum, from infrared to ultraviolet: Its band gap of about 1~eV~\cite{Ramos2021, Haines2018, Brec1979, FOOT1980189} makes 
FePS$_3$ suitable for infra-red detection while applications in ultraviolet photodetectors \cite{Gao_2018} and in non-linear optics\cite{xu2020controllable} have been reported as well. 
In addition, pure bulk FePS$_{3}$ displays antiferromagnetic ordering below 120 K~\cite{Lancon2016, lee2016ising} while its magnetic and structural properties can be tuned easily~\cite{Cheng2021} by chemical modification. 
That's why this material, similar to the low-temperature 2D magnet CrI$_3$, is a suitable candidate for the next step in spintronics towards AFM 2D spintronic devices~\cite{Ahn2020} operating at low temperatures.
%%%%%%%%%%%%%%%%%%%%%%%%%%%%%%%%%%%%%%%%%%%%%%%%%%%%%%%%%%%%%%%%%%%%%%%%%%%%%%%%%%%%%%%%%%%%%%%%%%%%%%%%%%%%%%%%%%%%%%%%%%%55
\begin{figure}[!htp]
   \centering
       \includegraphics[width=0.40\textwidth]{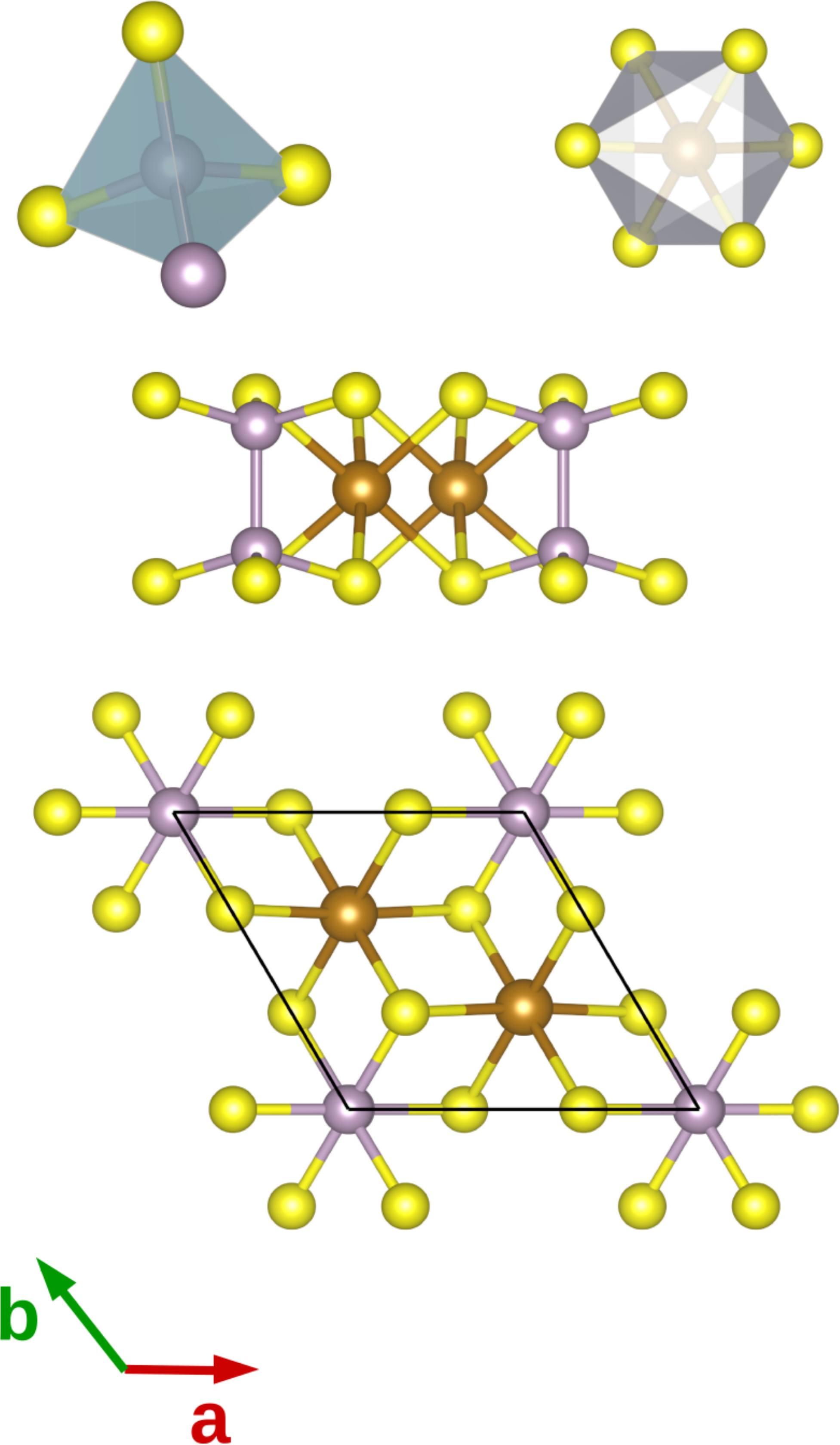}
        \caption{(Color online) (\textbf{Top row}) Octahedral and tetrahedral environment of Fe and P ions in FePS$_3$, respectively. The primitive cell of the FePS$_{3}$ monolayer (\textbf{Middle}) in side view and (\textbf{Bottom}) in top view. The brown, grey and yellow spheres denote the Fe, P and S atoms, respectively. Each Fe atom is surrounded by six S ions. The P dimer, oriented perpendicular to the plane of view, sits in the center of the Fe honeycomb. Magnetic exchange interactions between Fe ions are governed by the S ions as the mediating links between them. 
        Data in this figure, as well as in figures 3, 4 and 7, were drawn using the VESTA software \cite{vesta}.
        }
  \label{fig:geometry2}
\end{figure}
%%%%%%%%%%%%%%%%%%%%%%%%%%%%%%%%%%%%%%%%%%%%%%%%%%%%%%%%%%%%%%%%%%%%%%%%%%%%%%%%%%%%%%%%%%%%%%%%%%%%%%%%%%%%%%%%%%%%%%
\begin{figure}[!htp]
    \centering
   \includegraphics[width=0.8\columnwidth]{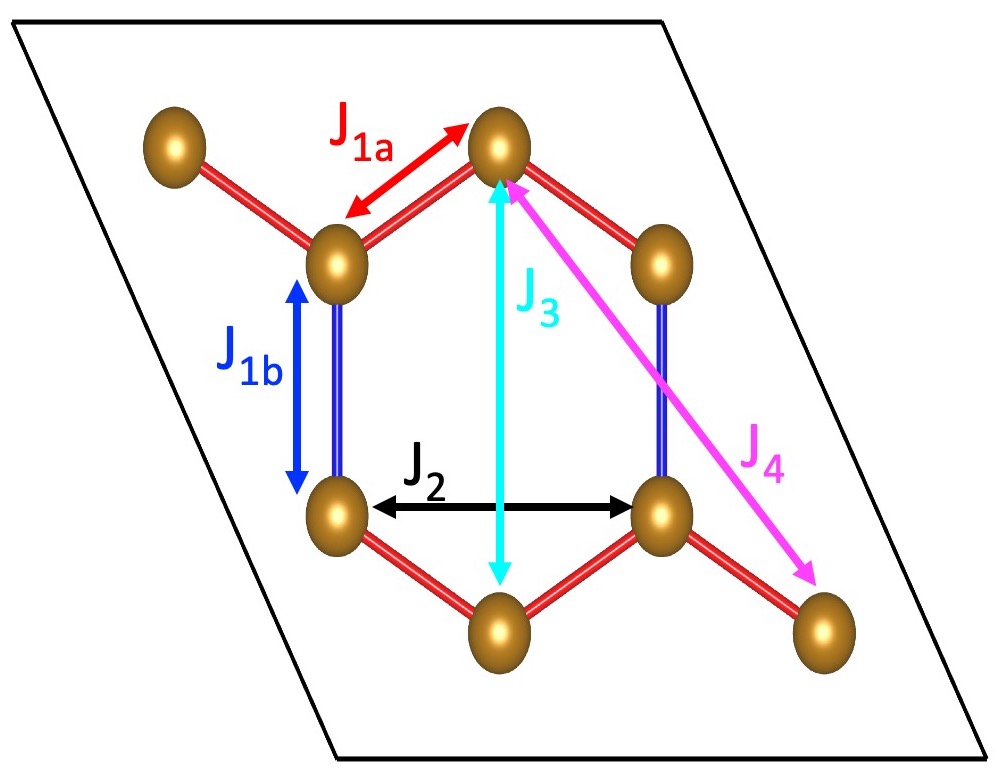}
   \caption{(Color online) Simplified representation of the FePS$_3$ crystal structure showing only the Fe atoms.
   A $(2 \times 2)$ cell is depicted, allowing us to show the (slightly) distorted honeycomb geometry in which Fe ions sit at each vertex. 
    The red and blue color marks short and long bonds, respectively, differing by about 4\% in length. 
   The magnetic exchange constants $J_i$ are indicated by the double arrows.
   }
   \label{fig:geometry1}
\end{figure}
%%%%%##################################################################################################################

%%%%%##################################################################################################################
%%%%%%%%%%%%%%%%%%%%%%%%%%%%%%%%%%%%%%%%%%%%%%%%%%%%%%%%%%%%%%%%%%%%%%%%%%%%%%%%%%%%%%%%%%%%%%%%%%%%%%%%%%%%%%%%%%%%%%
\begin{figure*}[tbp]
    \centering
    \includegraphics[width=1.3\columnwidth]{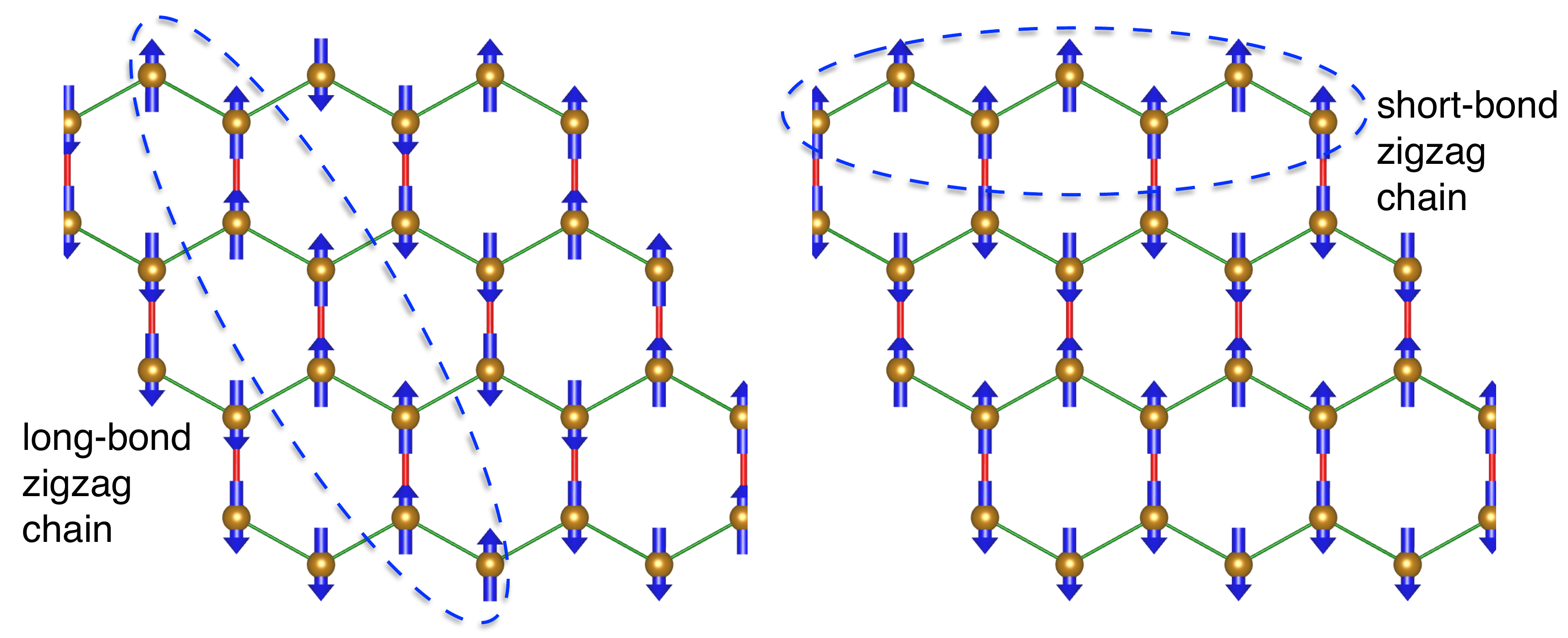}
    \caption{(Color online) The sublattice of Fe atoms in FePS$_3$ forms a distorted honeycomb structure. The blue arrows indicate the orientation of the Fe magnetic moments. 
   The Fe -- Fe bonds marked by red (thicker) lines are about 4\% longer than those marked by green (thinner) lines. 
    The (\textbf{left}) long-bond zigzag chain and (\textbf{right}) short-bond zigzag chain encircled by the blue, dashed ellipses 
    propagate along the crystallographic $b$-axis or $a$-axis, respectively. 
    }
   \label{fig:SLz}
\end{figure*}
%%%%%##################################################################################################################

From the structural point of view, bulk FePS$_{3}$ is a 2D layered material. 
Fig.~\ref{fig:geometry2} shows the unit cell of one monolayer. 
To visualize the crystal structure, it is better to write Fe$_{2}$P$_{2}$S$_{6}$, which means that in each layer, every Fe ion is surrounded by bipyramidal (P$_{2}$S$_{6}$)$^{4-}$ anions. 
In this way,  each Fe is octahedrally bonded to six S atoms while each P is bonded to three S atoms and one P atom.

The Fe$^{2+}$ ions in FePS$_3$ have the maximum magnetic moment compatible with their charge state, i.e. their spin is $S=2$.
Leaving away all the other ions for clarity, the positions of the Fe$^{2+}$ ions, and thus of their spins,  can be described by a distorted honeycomb lattice. 
Two opposite sides of  each hexagon are  slightly (between 0.017{\AA} and 0.19{\AA} according to different experimental sources~\cite{Klingen1973,Lancon2016}) longer than the other four sides, see Fig.~\ref{fig:geometry1}. This difference has a crucial effect on the type of magnetic exchange interaction between Fe atoms. 
Another important feature of the magnetism in FePS$_3$ is its strong out-of-plane easy axis, which renders stable magnetic order even in the monolayer limit. According to the Hohenberg-Mermin-Wagner theorem~\cite{Mermin1966, Hohenberg1967} in a system with isotropic interactions, thermal fluctuations in 2D would prevent the spontaneous symmetry breaking required to form a low-temperature ordered state. 
Therefore, some anisotropic interaction, e.g. magnetic single-ion anisotropy, is required in 2D systems to get a ground state with long-range magnetic order~\cite{Huang2017}.  However, despite many experimental and theoretical investigations~\cite{Olsen, Chittari, Lancon2016, Kurosawa, le1982magnetic, ZHANG2021167687, Wildes_2012, Rule2007}, the 
magnetic unit cell of the ground state of FePS$_{3}$ is still under debate.
Leaving aside possible complications in the three-dimensional material related to stacking order, several authors \cite{Kurosawa,Lancon2016} assumed an identity of the structural and the magnetic unit cell, containing two Fe$^{2+}$ ions each. Consequently, all Fe -- Fe bonds along a ferromagnetically aligned chain would be identical in length. 
In contrast, LeFlem {\it et al.} \cite{le1982magnetic} proposed a unit cell with four Fe$^{2+}$ ions in which the ferromagnetically aligned Fe chains run along an alternating sequence of short and long Fe -- Fe bonds.  This proposal was taken up later by other researchers~\cite{wildes2020, Murayama2016, OUVRARD19851181}, e.g. to explain the observation of Brillouin-zone folding effects in measured Raman spectra \cite{wang2016raman}. 
Note that the FM chains in both models are rotated by 60$^\circ$ in the plane, see Fig.~\ref{fig:SLz}.  

The so far unresolved issue of the relation between lattice distortion and magnetic ground state 
motivated us to carry out a comprehensive computational study of the magnetic 
properties employing the DFT+$U$ approach (Density Functional Theory plus on-site electron-electron repulsion). 
In the literature, spin model Hamiltonians had considered exchange interactions only up to the third-nearest neighbors of the Fe atoms \cite{Chittari, Olsen}, and the difference between short and long Fe -- Fe bonds had been ignored in the Hamiltonian parameterization. 
Moreover, higher-order (in the spin variable) couplings, such as the Dzyaloshinskii-Moriya interaction and biquadratic couplings, had been neglected so far.  
We find that $U=$2.22~eV can produce the experimental band gap of 1.23~eV.  Interestingly, we find that the magnetic exchange interaction is ferromagnetic and antiferromagnetic along the long bonds and short bonds, respectively. This assignment turns out to be robust with respect to changes in the $U$ parameter. We conclude that the magnetic ground state consists of ferromagnetic spin chains running along the long bonds that couple antiferromagnetically among each other. 
In addition, we propose a model spin Hamiltonian that includes anisotropic spin interactions that are of crucial importance of 2D magnets, as well as an interaction up to forth neighbors, such that the fall-off of magnetic interactions with distance, as it is expected for a magnetic insulator, can be seen.
%%%%%%%%%%%%%%%%%%%%%%%%%%%%%%%%%%%%%%%%%%%%%%%%%%%%%%%%%%%%%%%%%%%%%%%%%%%%%%%%%%%%%%%%%%%%%%%%%%%%%%%%%%%%%%%%%%%%%%
\begin{figure*}[!htp]
    \centering
    \includegraphics[width=0.65\columnwidth]{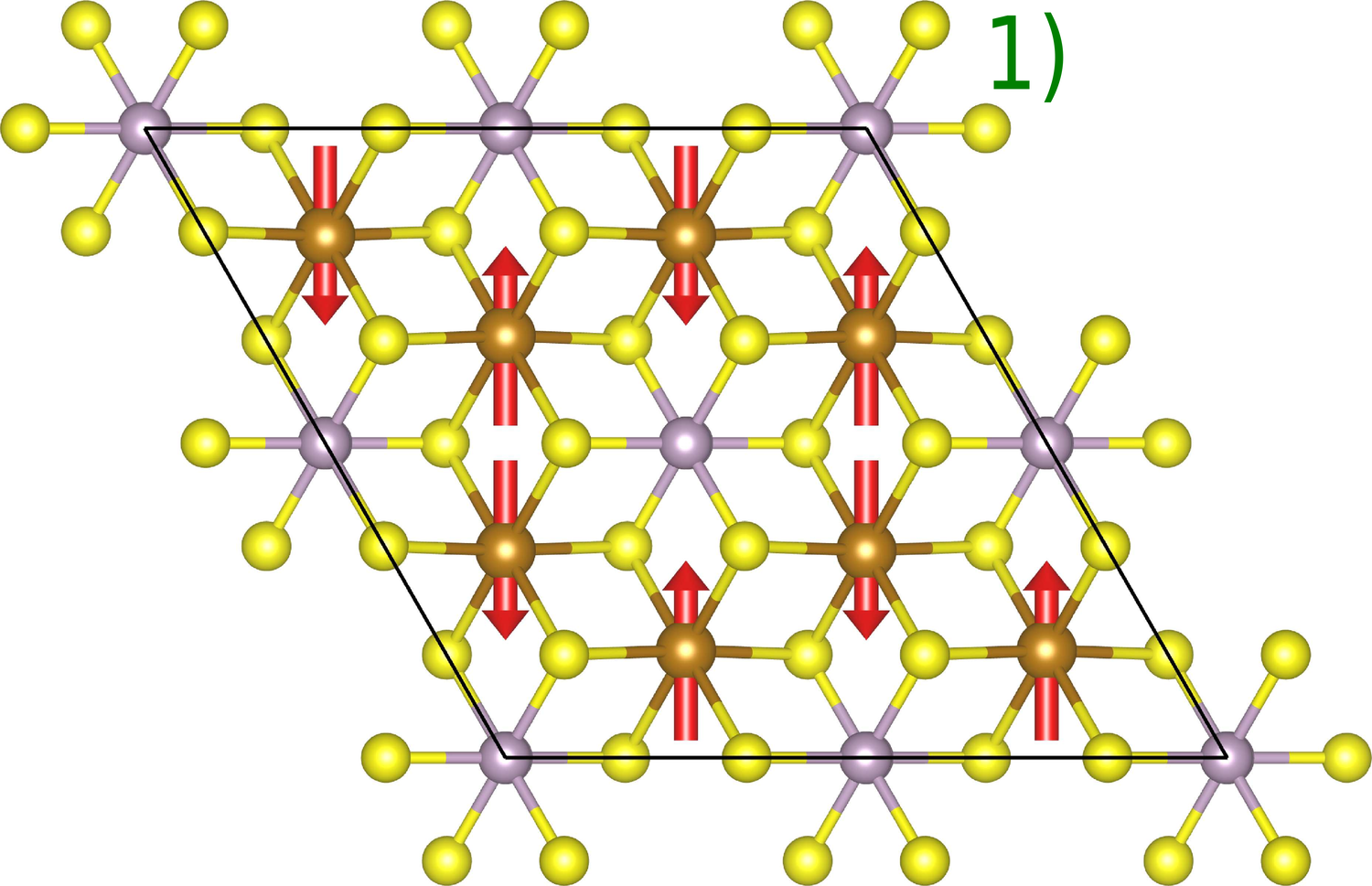}
        \vspace{0.5cm}
    \includegraphics[width=0.65\columnwidth]{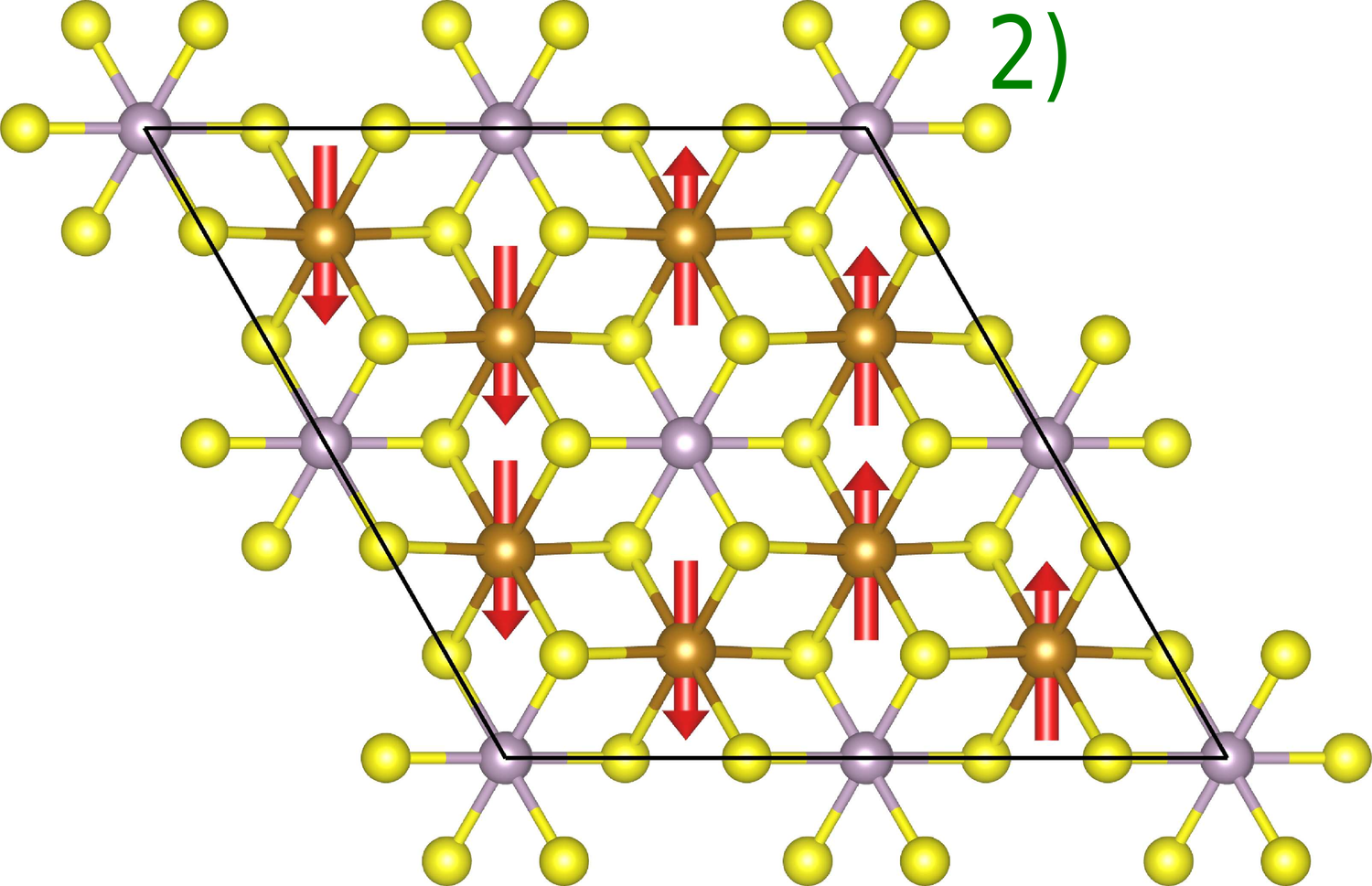}
        \vspace{0.5cm}
    \includegraphics[width=0.65\columnwidth]{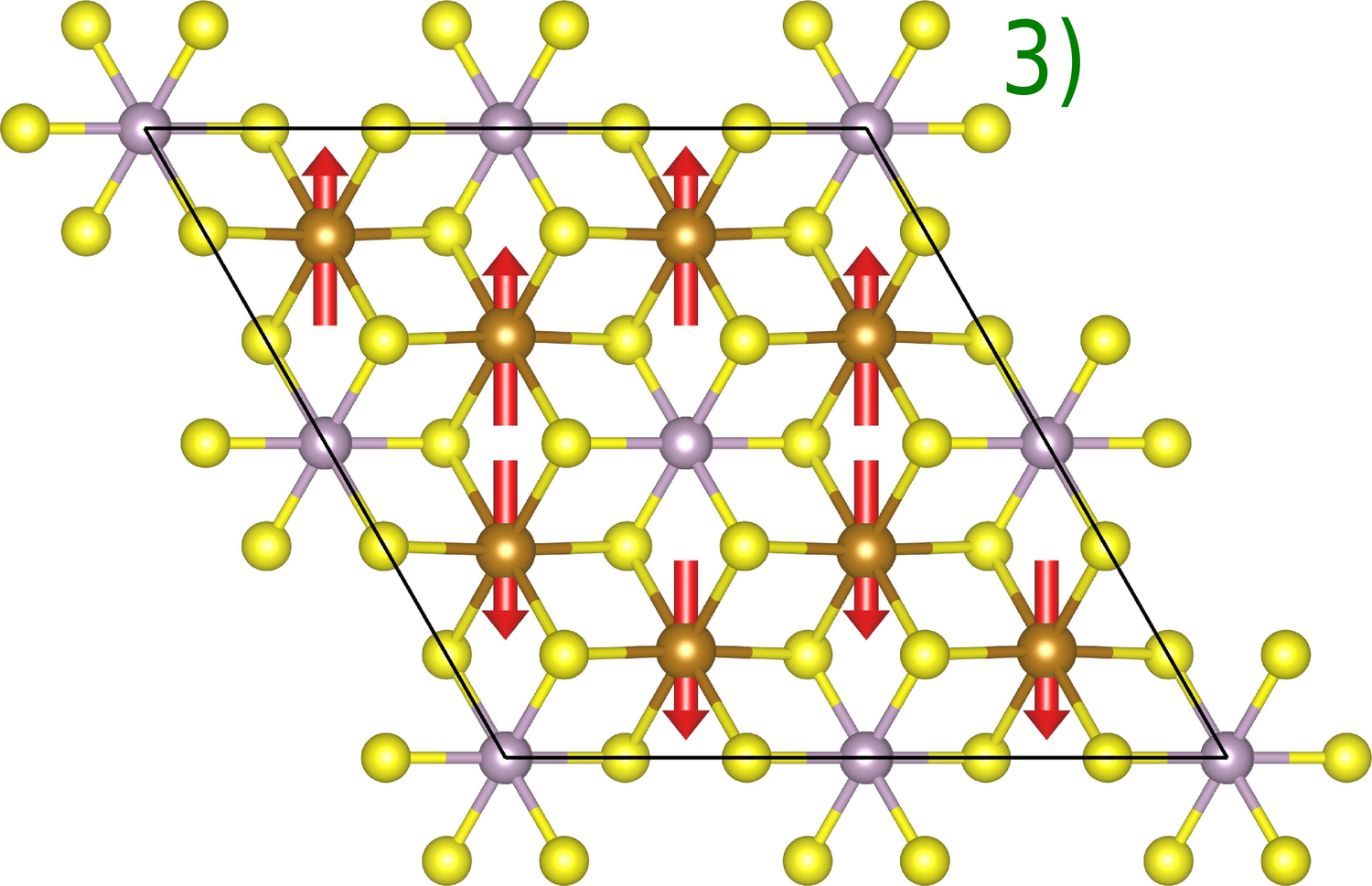}
        \vspace{0.5cm}
     \includegraphics[width=0.65\columnwidth]{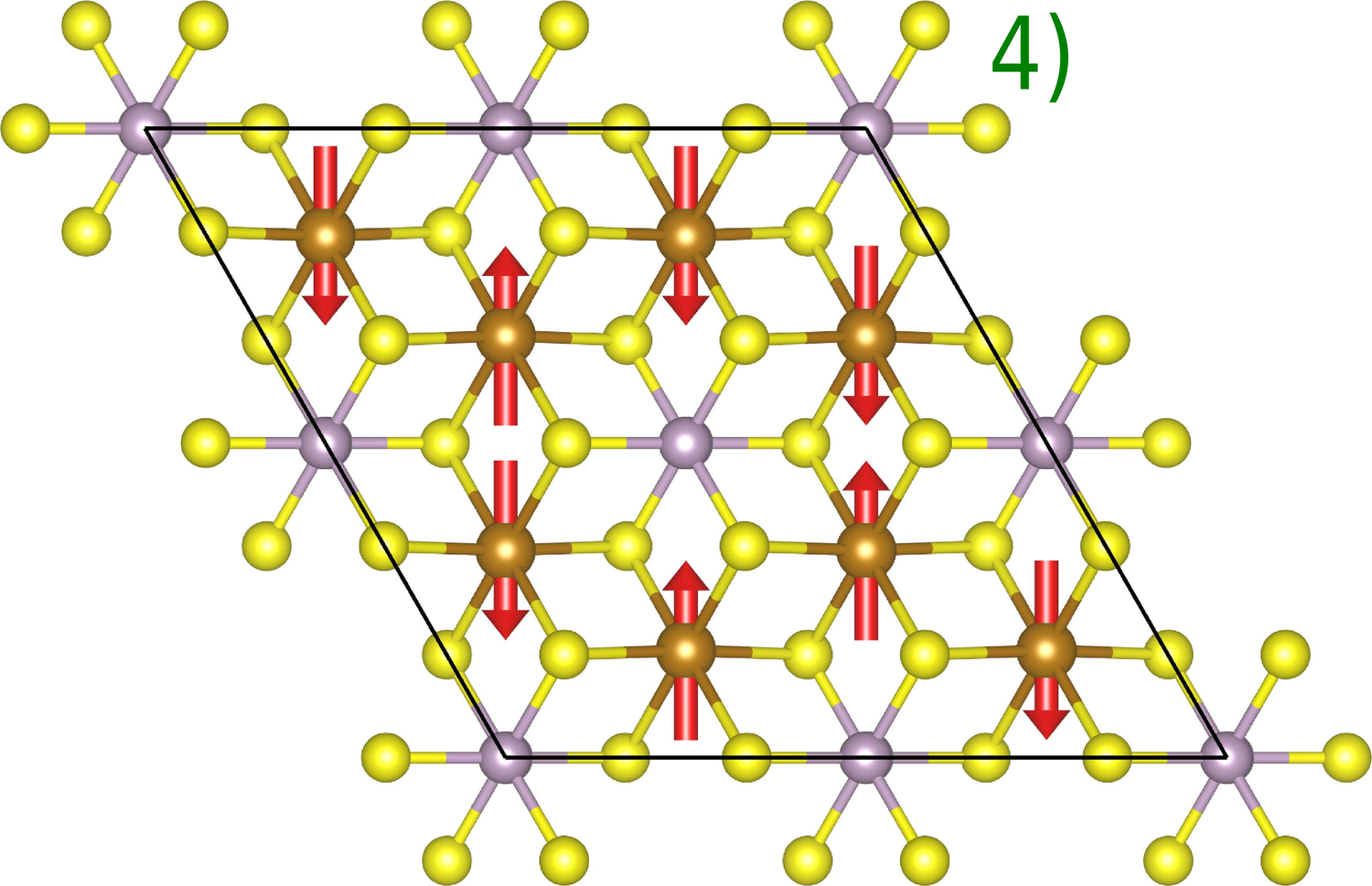}
      \includegraphics[width=0.65\columnwidth]{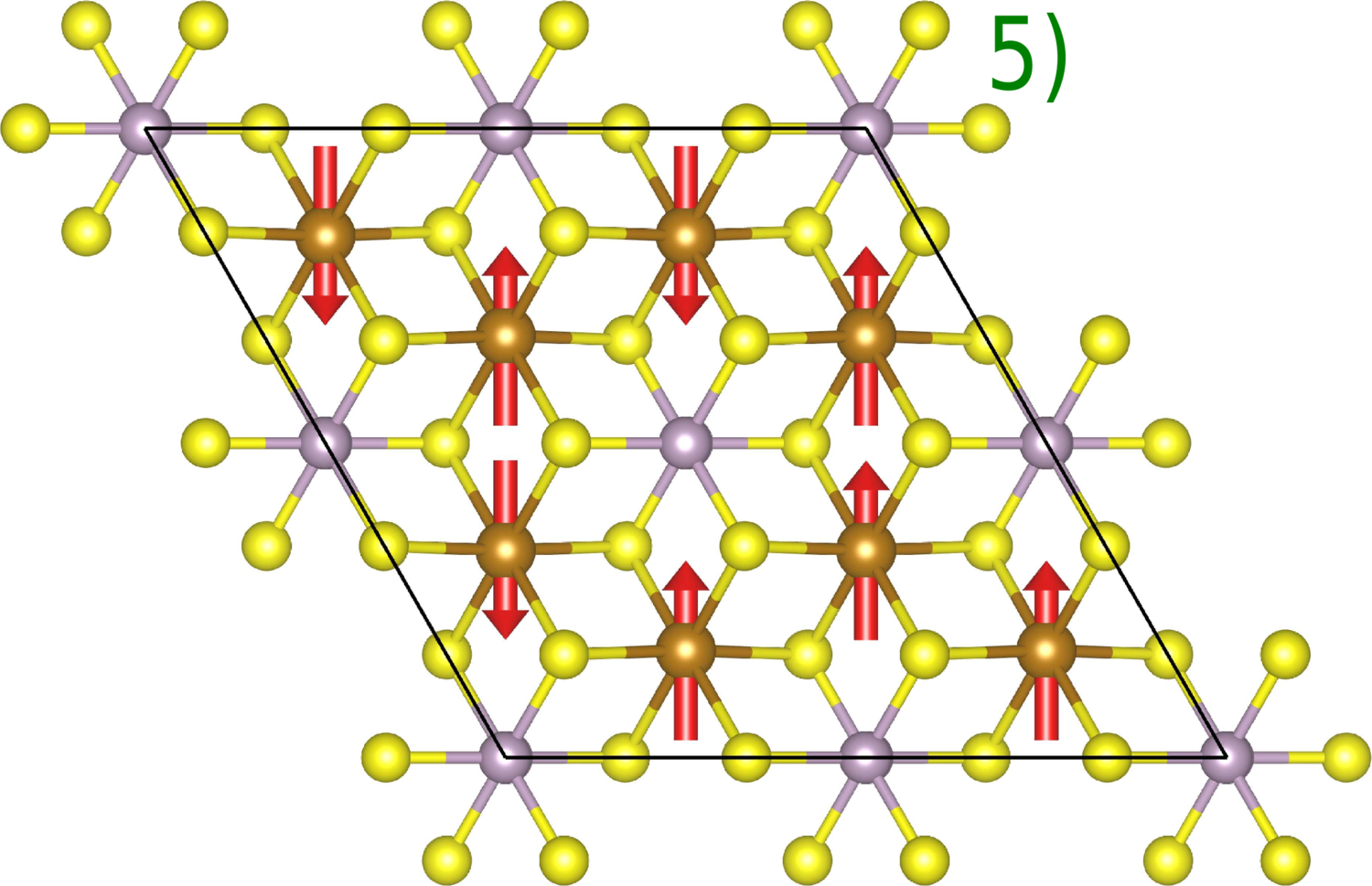}
       \includegraphics[width=0.65\columnwidth]{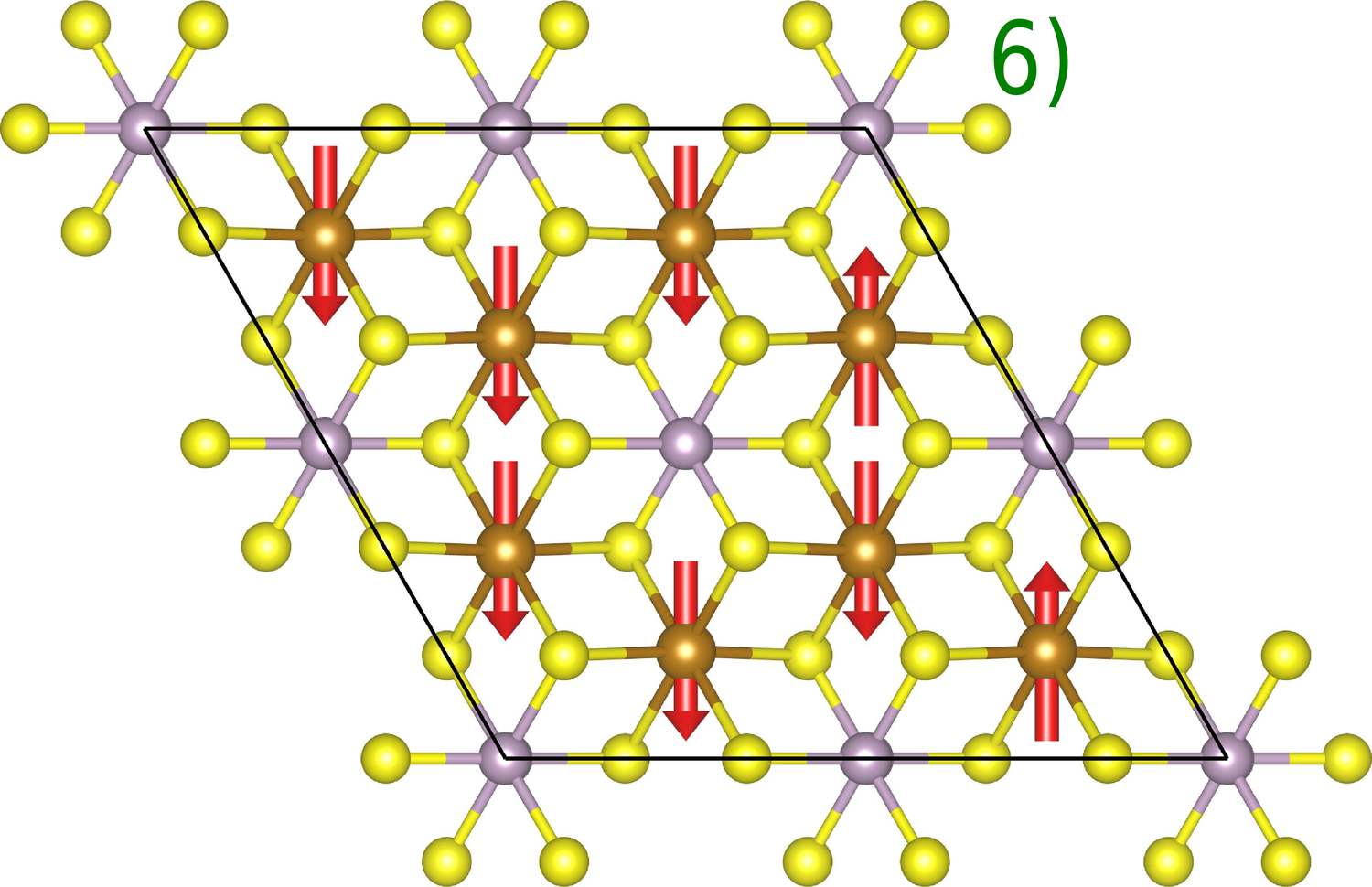}
   \caption{(Color online) Different collinear magnetic configurations that we use for obtaining $J$ parameters. 
   The yellow, grey and brown spheres are S, P and Fe ions, respectively. The red arrow shows the direction of the magnetic moment on each Fe ion. 
   Note that some Fe-- Fe distances (the vertical ones in the plots) are about 4\% larger than the others, and hence all six configurations shown are non-equivalent. 
   \textbf {Top row} shows (from left to right) Neel, long-bond zigzag and short-bond zigzag configurations. Our DFT+$U$ calculations show that the long-bond zigzag is the global minimum ground state. 
   \textbf{Bottom row} shows additional collinear configurations considered.
   }
   \label{fig:Hzn}
\end{figure*}
%%%%%##################################################################################################################
%%%%%%%%%%%%%%%%%%%%%%%%%%%%%%%%%%%%%%%%%%%%%%%%%%%%%%%%%%%%%%%%%%%%%%%%%%%%%%%%%%%%%%%%%%%%%%%%%%%%%%%%%%%%%%%%%%%%%%%
\begin{figure}[!htp]
    \centering
    \includegraphics[width=1.0\columnwidth]{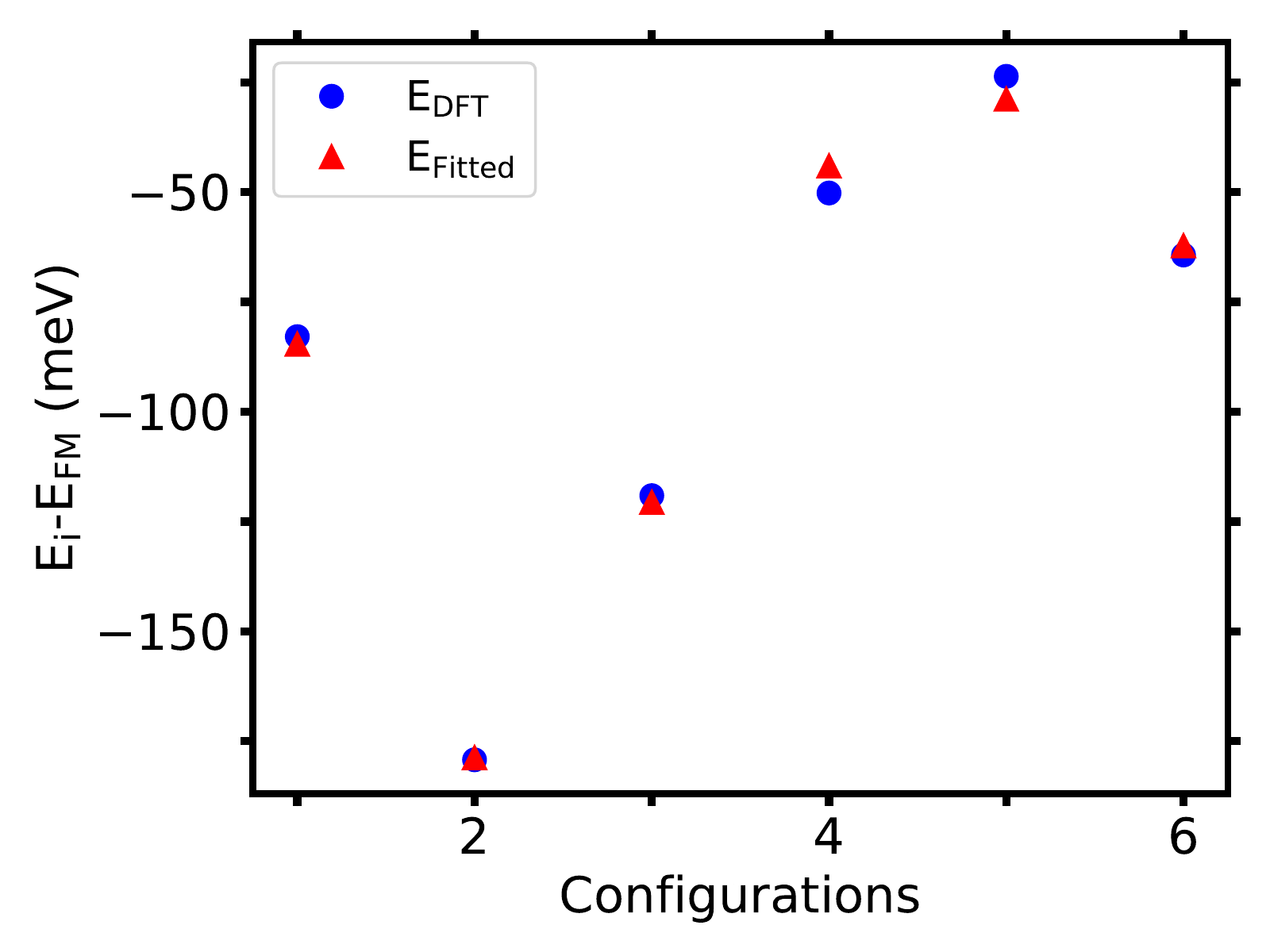}
   \caption{(Color online) This figure illustrates how total energies of different collinear configurations obtained by means of DFT+$U$ calculations are mapped to a Heisenberg Hamiltonian~\cite{li2021spin, Sadeghi2015}. With this mapping, we can calculate the $J$ parameters for various neighbor shells. The numbered spin configurations correspond to those shown in Fig.~\ref{fig:Hzn}. 
   The energies plotted in the graph are for $U_\text{eff}$=2.22 eV. The standard fitting error for each $U_\text{eff}$ is $\sim 2$}\%.
   \label{fig:fit1}
\end{figure}
%%%%%##################################################################################################################

The paper is structured as follows. In section~\ref{sec:C-details} the details of the DFT and Monte Carlo calculations are presented.
Section~\ref{sec:R&D} is devoted to electronic properties as well as the derivation of the spin Hamiltonian and the different aspects of exchange interactions in the determination of the magnetic ground state of the FePS$_{3}$ monolayer.
Finally, In section~\ref{sec:Conclusion} a summary is given.

\section{Computational details}
\label{sec:C-details}

As a single crystal, iron phosphorous trichalcogenide  crystallizes in the monoclinic structure with space group C2/m. The Fe atoms occupy the 4g(0,y,0) sites, P atoms occupy the 4i(x,0,z) and S atoms occupy 4i(x,0,z) and 8j(x,y,z) sites, respectively. 
Structural parameters can be found in Ref.~\onlinecite{Lancon2016}.
We used the so-defined lattice parameters and atomic positions as starting point for our calculations of a FePS$_3$ monolayer. To generate an isolated layer, the lattice parameter perpendicular to the layers ($c$-axis) of the bulk structure was increased to 20~\AA.
 %%%%%%%%%%%%%%%%%%%%%%%%%%%%%%%%%%%%%%%%%%%%%%%%%%%%%%%%%%%%%%%%%%%%%%%%%%%%%%%%%%%%%%%%%%%
 \begin{table}[bth]
   \centering
   \caption{Optimized nearest-neighbor distances obtained from 
        DFT+$U$ calculations using $U=2.22$~eV of the long-bond zigzag ground state 
        compared to distances determined experimentally from Ref.~\onlinecite{Lancon2016}. 
        For the derivation of the spin Hamiltonian, the different lengths of nearest-neighbor bonds $d_\text{1a}$ and $d_\text{1b}$ are considered by different exchange interactions $J_\text{1a}$ and $J_\text{1b}$ 
        while the same exchange constant is used for all atoms in the second to forth neighbor shell.
   }
       \begin{tabular}{l|c|c}
                                    & this work  & experimental \cite{Lancon2016} \\
      \hline
     $d_\text{1a}$ (\AA) &  3.44  & 3.36  \\
     $d_\text{1b}$ (\AA) &  3.58  &  3.55 \\
     $d_\text{2}$ (\AA)  &    6.02, 6.05  & 5.93, 5.94 \\ 
     $d_\text{3}$ (\AA) &   6.92, 7.00 & 6.71, 6.92 \\
     $d_\text{4}$ (\AA) &  9.17, 9.18, 9.33 & 8.96, 9.04, 9.17 \\
     \hline
       \end{tabular}
        \label{tab1}
 \end{table}
 %%%%%%%%%%%%%%%%%%%%%%%%%%%%%%%%%%%%%%%%%%%%%%%%%%%%%%%%%%%%%%%%%%%%%%%%%%%%%%%%%%%%%%%%%%%%%%%%%%%%%%%%%%%%%%%%%%%%%%%

We employ two different computational approaches for our first-principles calculations: 
For calculating the single-site magnetic anisotropy and anisotropic interactions between spins that rely on a relativistic description of electrons as well as for the biquadratic interaction which needs a non-collinear scheme, we use an all-electron full-potential linearized augmented plane-wave (FPLAW) method.  
For large systems consisting of many atoms with collinear spins, 
computationally more efficient 
calculations were carried out with the Quantum-Espresso (QE)~\cite{Giannozzi_2009} code in the framework of the spin-polarized density functional theory.
We approximate the exchange-correlation energy using the generalized gradient approximation (GGA) in the Perdew-Burke-Ernzerhof parameterization~\cite{Perdew1996}. 
For a better description of the low-temperatures ground state of FePS$_{3}$, we used GGA+$U$ approach to correct on-site electron-electron interaction ($U$) for the $3d$ orbitals of the Fe atoms, following Dudarev's approach which includes a spherically symmetric effective on-site Coulomb repulsion $U_\text{eff}$. 
Since QE is a plane-wave code,  GBRV ultra-soft pseudo-potentials~\cite{Vanderbilt2014} for the elements Fe, P, and S are employed to describe the interaction of the valence electrons with the ionic core. 
The optimized cut-off energies of 50 Ry and 480 Ry have been used for expanding the wavefunctions and charge density, respectively, in plane waves.

Some magnetic modelling parameters, such as the single-ion anisotropy (SIA) and the Dzyaloshinskii-Moriya (DMI), require the inclusion of spin-orbit interaction, and therefore a relativistic treatment of the electrons. 
For an accurate treatment of these aspects, we employ the full-potential linearized augmented plane wave (FPLAW) method, as implemented in the FLEUR code~\cite{fleur}. 
The cut-off energy of the plane-wave expansion in the interstitial region is set to $k_{\mathrm{max}}=3.8\, \mathrm{a.u.}^{-1}$. The muffin-tin radii of Fe, P and S atoms are set to 2.8, 1.49 and 1.90 a.u., respectively.
Although the FLEUR code allows for a more general treatment of the on-site electron-electron interaction, we chose to set the on-site Hund exchange to zero to stay compatible the the QE calculations. 

In order to determine the type of magnetic order of the ground state, we define a model spin Hamiltonian :
\begin{align}
H_{\rm {spin}} & = H_{\rm Heis}     +\frac{1}{2} B\sum_{\rm n.n} (\vec{S_{i}}\cdot\vec{S_{j}})^{2} \nonumber \\
  & +\frac{1}{2} D \sum_{\rm n.n} \hat{D}_{ij}\cdot(\vec{S_{i}}\times \vec{S_{j}})+\Delta\sum_{i} (\vec{S_{i}}\cdot\vec{d_{i}})^{2}
\label{H}
\end{align}
where $\vec{S_\text{i}}$ represents direction of magnetic spins, $H_{\rm Heis}$ is the usual Heisenberg Hamiltonian (for details see below), 
$B$, $D$ and $\Delta$ are the strengths of biquadratic, DMI and SIA, respectively. 
Moreover, unit vectors $\hat{D}_\text{ij}$ and $\vec{d_\text{i}}$ show the direction of the DMI  and the easy axis of magnetization at each site $i$, respectively. 
It should be noted that the direction of DMI is determined by Moriya rules~\cite{Moriya1960}. Due to the centrosymmetric 2/m point group symmetry,  the FePS$_{3}$ monolayer has a mirror plane perpendicular to the $b$-axis. 
According to the Moriya rules, when a mirror plane includes two ions, the $D$ vector should be perpendicular to the mirror plane.   

%%%%%%%%%%%%%%%%%%%%%%%%%%%%%%%%%%%%%%%%%%%%%%%%%%%%%%%%%%%%%%%%%%%%%%%%%%%%%%%%%%%%%%%%%%%%%%%%%%%%%%%%%%%%%%%%%%%%%%%
\begin{figure*}[!htp]
    \centering
    \includegraphics[width=1.0\columnwidth]{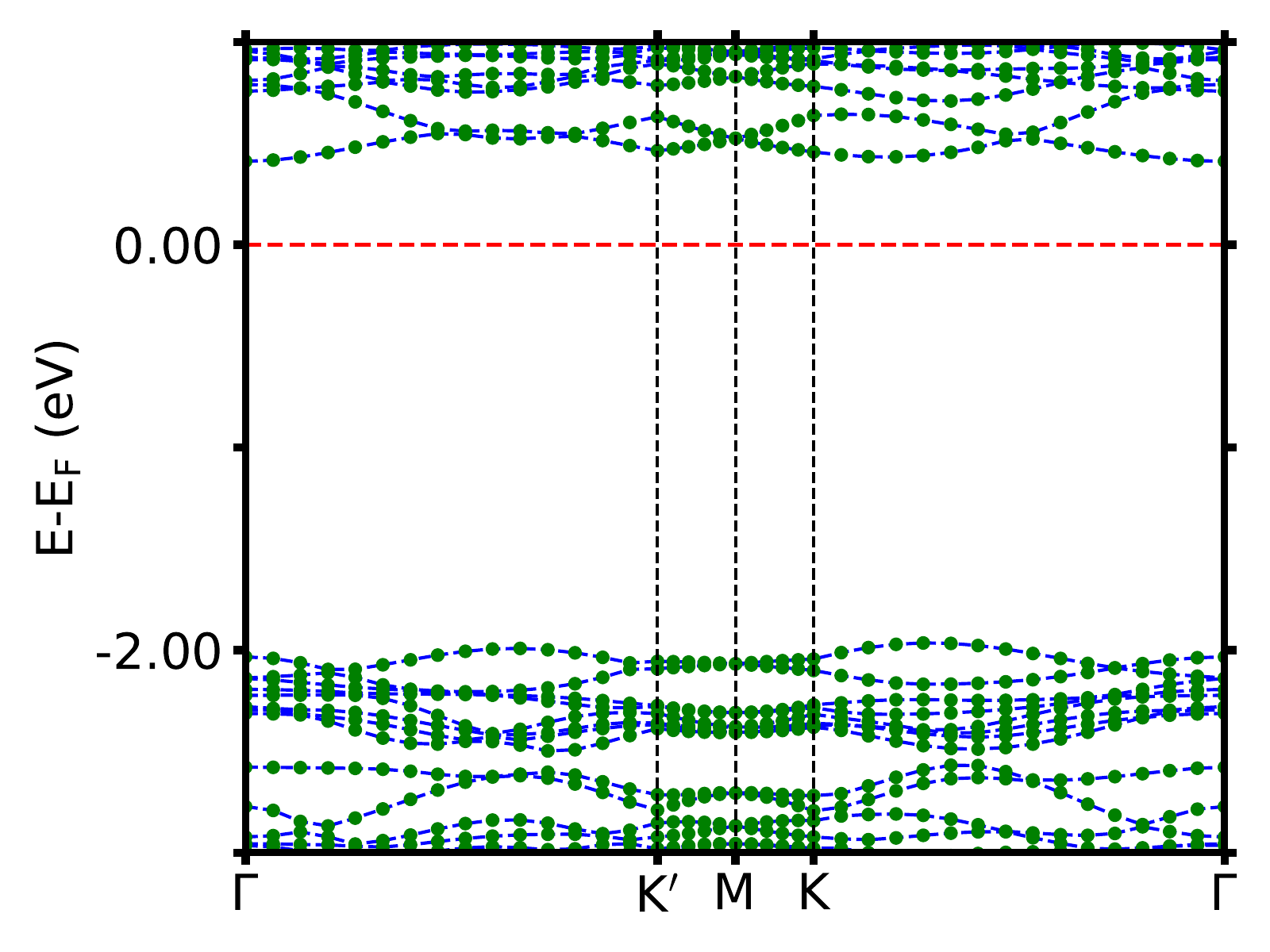}
    \includegraphics[width=1.0\columnwidth]{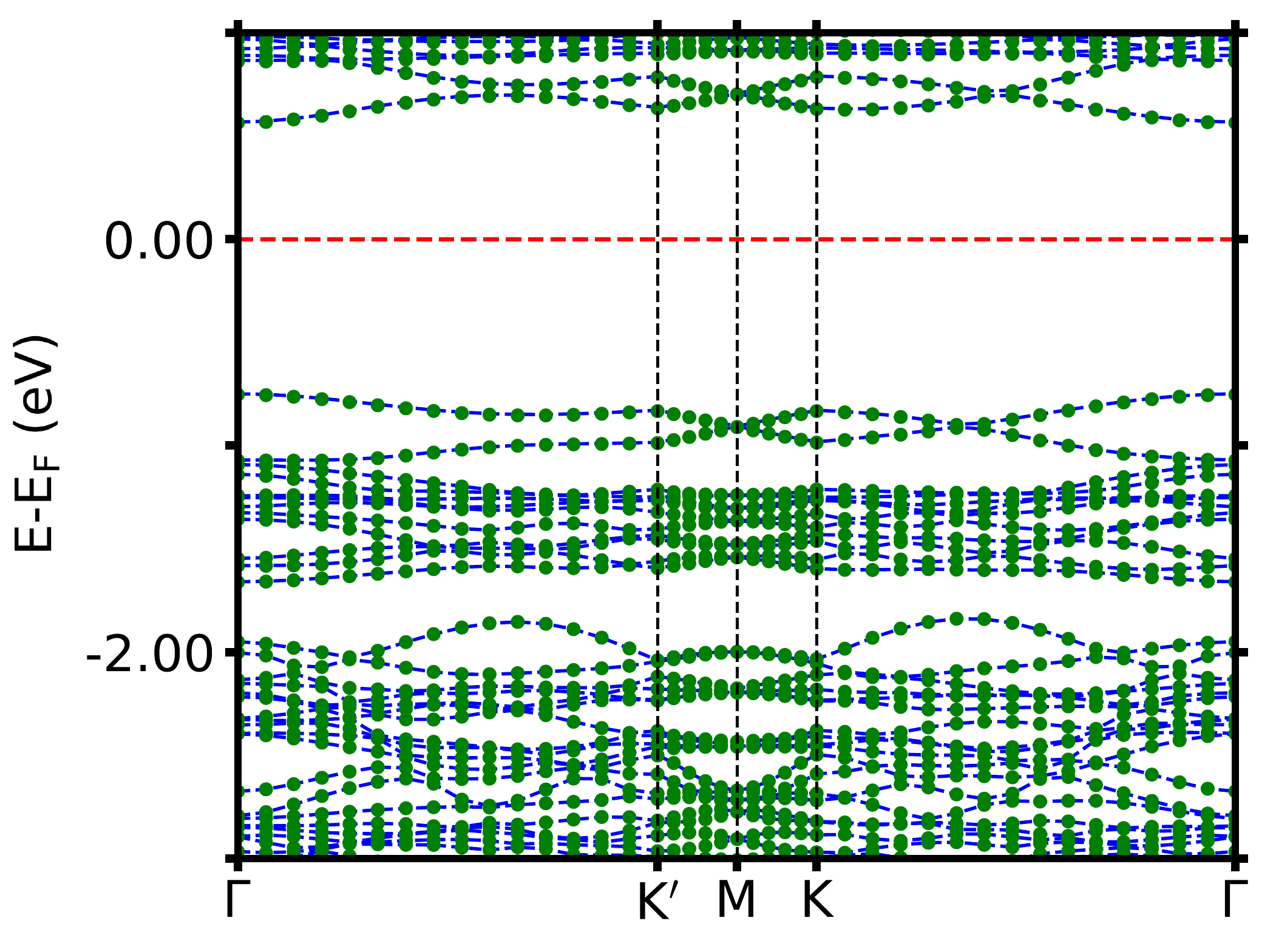}
        \includegraphics[width=1.0\columnwidth]{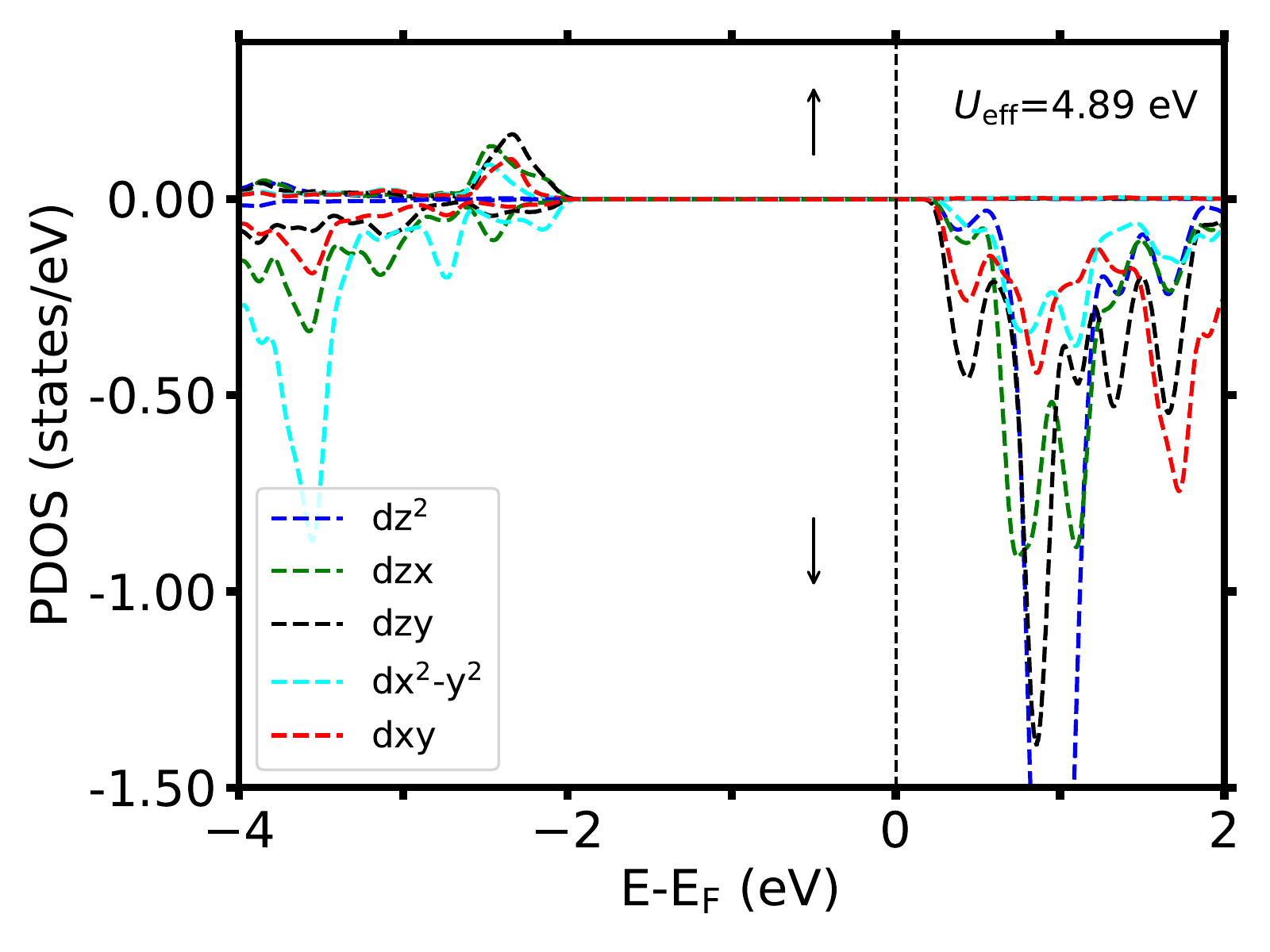}
    \includegraphics[width=1.0\columnwidth]{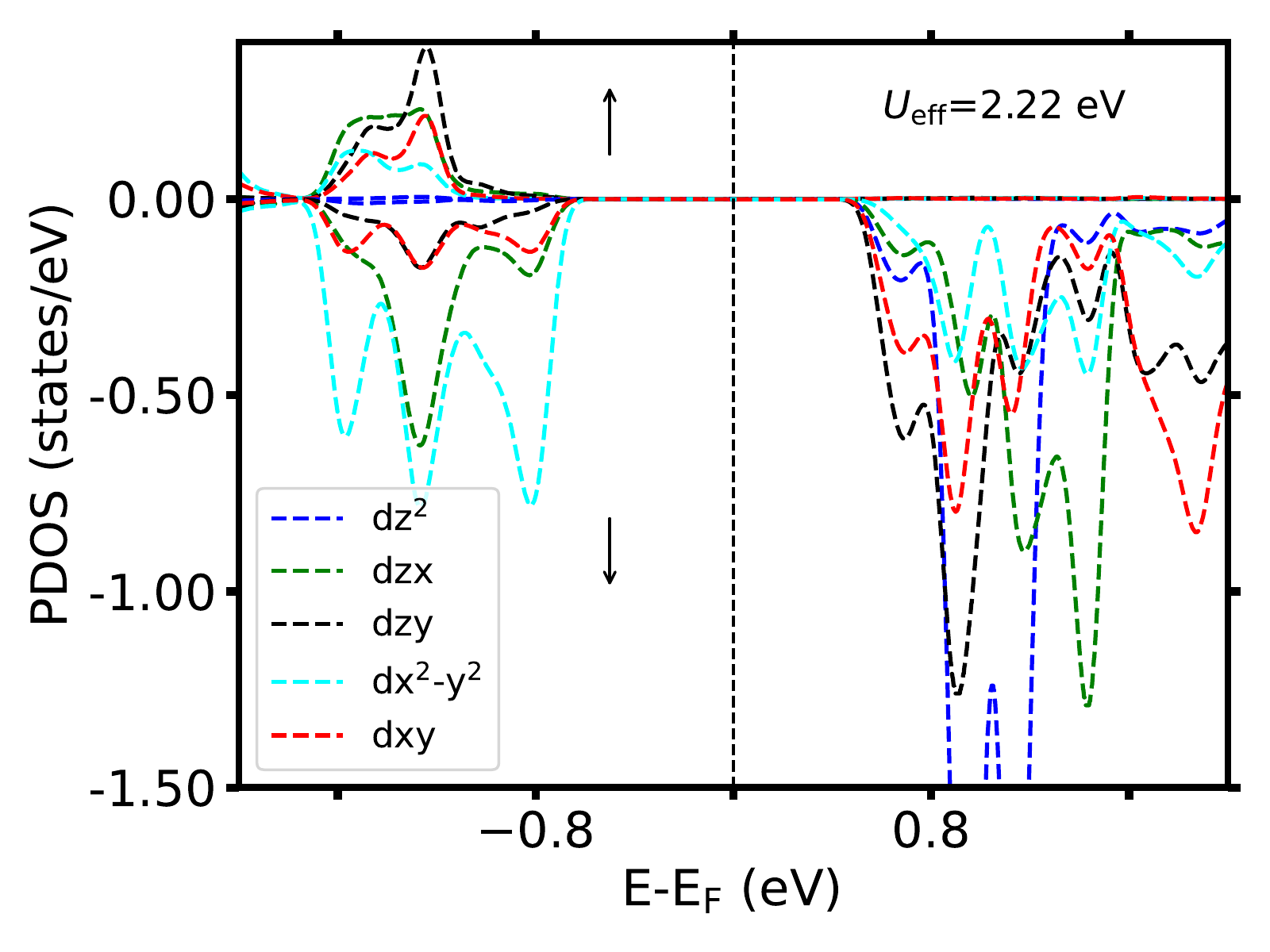}
   \caption{(Color online) \textbf{Top row} Bandstructure  and (\textbf{Bottom row}) projected density of states (PDOS) of the AFM long-bond zigzag structure for $U_{\textrm{eff}}=$~4.89 and 2.22~eV, respectively. For bandstructures, the green and blue colors denote spin up and spin down, respectively. Due to the absolute spin being zero for AFM long-bond zigzag, the two spin channels collapse to each other.  When $U_{\textrm{eff}}$ is decreased, the contribution of the $d_{x^{2}-y^{2}}$ orbital to the highest occupied bands increases. This is the reason why in the  bandstructure with $U_{\textrm{eff}}=2.22$~eV the two topmost valence bands are split off. 
    These calculations were done in the $(2 \times 2)$ cell also used for the calculations of the $J$ parameters.
    }
   \label{fig:elect}
\end{figure*}
%%%%%##################################################################################################################
For the determination of $B$, $D$ and $\Delta$ with FLEUR, we consider the primitive cell (2~Fe atoms, 10 atoms in total) and a $10\times10\times1$ Monkhorst-Pack k-point mesh. 
The biquadratic interaction originates from electron hopping, and 
in terms of a tight-binding description it can be obtained in fourth order from a  perturbation theory~\cite{Mila2000, Tanaka2018}. 
For extracting the $B$ term from DFT calculations, we rotate the spins of the two Fe atoms pointing in opposite direction around the $c$-axis in such a way that 
the sum of them remains zero, $S_{1}+S_{2}=$0. In this way, the energy due to the Heisenberg term is degenerate 
and the total energy differences are attributed to $B$ only. 
It should be noted that in the calculation of $B$ spin-orbit coupling (SOC) has not been considered, and the anisotropic terms are quenched.   
For the anisotropic exchange interactions, we need to consider the effect of SOC (GGA+$U$+SOC).  
Care must be taken in the selection of appropriate spin configurations that allow us to extract the values of $D$ and $\Delta$ from total energy differences. 
Therefore, pairs of configurations are constructed in such a way that their energies, as given by the Heisenberg term, are degenerate before turning on SOC.
For SIA, we consider a pair of magnetic configurations in which all spins are oriented along the $c$-axis or along the $b$-axis, respectively. 
For these configurations, the DMI term vanishes in GGA+$U$+SOC calculations. For obtaining the strength of $D$, we need to consider two different magnetic configurations in which two spins are oriented on the $a$-axis and the negative $b$-axis for the first, and the $a$-axis and the positive $b$-axis for the second one. In this way, the $\Delta$ term vanishes in GGA+$U$+SOC calculations. 
Form energy difference of each pair of configurations, the $\Delta$ and $D$ can be extracted.

With the aim to characterize the low-temperature magnetic ground state of 
FePS$_{3}$, Monte Carlo
simulations of classical spins on a lattice have been performed using the replica-exchange method~\cite{Hukushima1996}.
We use two-dimensional lattices consisting
of $N \times L^{2}$ spins, where $L = 10$ is the linear size of the
simulation cell and $N$ is the number of spins  ($N = 2$ for the 
primitive cell).
For checking phase transition at low temperature, different sizes of the simulation cell from $L=8$ to $L=14$ are considered. For the thermal equilibrium
and data collection, we consider 3 $\times$ 10$^{6}$ Monte Carlo steps (MCs) per spin at each temperature.

\section{Results and discussion}
\label{sec:R&D}

\subsection{Geometric and electronic structure} 
\label{sec:G&E}

It is well known from previous studies that in FePS$_3$ the Fe$^{2+}$ ions are in the high-spin state of Fe. Consequently at one  Fe atom five electrons are pointing spin-up while one electron is spin-down and occupies a single $3d$ orbital. Of course, at Fe atoms with opposite magnetic moment the role of spin-up and spin-down electrons is interchanged. 
Due to the rather large distance between the Fe$^{2+}$ ions, the interaction between the $3d$ orbitals of different Fe ions is weak, yet decisive for the magnetic ground state.   
Since the energies of the orbitals on the $3d$ shell are split up by the crystal field of the surrounding sulfur anions, 
unequal occupation of the 3d orbitals, and hence orbital ordering, are to be expected. 
Thus, the DFT$+U$ method is mandatory for a correct description of electronic structure. 
Several approaches are possible to determine a suitable value for $U$: 
We found that we are able to reproduce the experimental band gap 
by choosing $U=2.22$ eV. For this purpose, we rely on the most recent experimental value 
of  1.23~eV\cite{Ramos2021}, discarding an older optical absorption measurement that had estimated a band gap of $\sim 1.6$~eV\cite{Brec1979} as outdated. This value of $U$ is of the same order as the value $U=2$eV used in Ref.~\onlinecite{Olsen}. 
Alternatively, we estimated $U$ using Density Functional Perturbation Theory (DFPT)~\cite{Timrov2018} which is implemented in the QE code.
This calculation has been done by considering the primitive cell of FePS$_{3}$ (containing 10 atoms), and a value of $U=4.89$eV is obtained.  Since the DFT$+U$ calculations have been done using two codes, we compared the obtained gap to make sure that different implementations of DFT+$U$ have no effect on the results. The gaps obtained from the QE and FLEUR code are in good agreement for each $U$ parameter.

We started by performing a number of explorative calculations to investigate the interplay between structure and magnetic order at low temperature.  
For this purpose we used a $(1 \times 2)$ cell which includes four Fe ions and allows us to build ferromagnetic zigzag chains of spins, either running along the long or the short Fe -- Fe separations, as illustrated in Fig.~\ref{fig:SLz}. 
We started from the experimental geometry~\cite{Lancon2016} and optimized the geometry using self-consistent forces obtained with the GGA+$U$ method. 
In particular, we optimized both the lattice vectors and the positions of the atoms in the cell for both spin configurations shown in Fig.~\ref{fig:SLz}. 
In both cases, it is found that the Fe -- Fe distances are unequal; two opposing edges of the Fe hexagons are longer than the others by 0.14~{\AA} and 0.13~{\AA} for 'long-bond' and 'short-bond' spin configurations, respectively. The lattice constants determined for the two spin structures differ by less than 0.4\%. 
Having spins ferromagnetically aligned in the zigzag chain passing through the long Fe--Fe bonds is found to be always  energetically more favorable 
than the alignment along the short bonds, or any other spin structure in this cell. 
Moreover, we find this trend to be independent on the value of $U$.  
To construct the spin Hamiltonian, it is therefore justified to work with the 
optimized lattice constants and atomic positions of the ground-state spin configuration, the long-bond zigzag chain, 
and to use this fixed geometry for all others. 
The optimized distances between the Fe atoms which are used to derive $J$ parameters are summarized in Table~\ref{tab1}.
Using $U = 2.22$eV, the two different nearest neighbor distances are determined as $d_{1a}=$3.44~{\AA} and $d_{1b}=$3.58~{\AA}, see  Fig.~\ref{fig:geometry1}.
Thus, the obtained results show that the distance between the nearest Fe neighbors differs by 
$d_{1b} - d_{1a} = 0.14${\AA}; this agrees well with the structure determination by neutron scattering at low temperature by Lan{\c{c}}on {\em et al.} \cite{Lancon2016} who obtained 
0.19 {\AA} while earlier experiments~\cite{Klingen1973} using X-ray diffraction at room temperature in magnetically disordered samples had found a smaller value of 0.017{\AA}.
We will later argue that this distortion has a crucial effect on exchange interaction and magnetic properties of the ground state. 
The obtained lattice parameters for the primitive unit cell ($a=6.017$~{\AA} and $b=6.052$~{\AA} and $\gamma$=119.86$^{\circ}$) are in good agreement with experimental values\cite{Lancon2016} of bulk samples ($a=5.940$~{\AA} and $b=5.972$~{\AA} and $\gamma$=121.29$^{\circ}$). 

Calculations of the electronic band structure and the orbital-projected density of states of the magnetic ground state for both values of $U$ are shown in Fig.~\ref{fig:elect}. 
While the band structure obtained with $U=2.22$eV reproduces the experimental gap of 1.23eV, the calculation with $U=4.89$eV gives a much wider band gap.  
The projected density of states show that for $U=2.22$eV this gap is opened up by the intra-atomic Coulomb repulsion between the occupied $3d_{x^2 - y^2}$ orbitals and the remaining unoccupied $3d$ orbitals. 
For the larger $U_\text{eff}$ value, the occupied $3d$ state of Fe is pushed down even further in energy and energetically overlaps with the valence bands, hence the larger band gap encountered in this case. 
A L\"owdin analysis indicates that all $3d$ states for spin majority are fully occupied 
while for spin minority the L\"owdin charges in the individual orbitals for $U_\text{eff}$=4.89 eV are  0.03,  0.35,  0.18,  0.60, 0.15 for  $d_{z^{2}}$, $d_{xz}$, $d_{yz}$,  $d_{x^{2}-y^{2}}$ and $d_{xy}$, respectively. 
By decreasing the $U_\text{eff}$ parameter, the hybridization between the $3d$ orbitals of Fe and the $p$ orbitals of P and S slightly  weaker, so that for  $U_\text{eff}$=2.22 eV the $d_{x^{2}-y^{2}}$ orbital has less mixing with other orbitals, and the respective L\"owdin charges are 
0.02, 0.31, 0.13, 0.61, and 0.15. 
We conclude that $U_\text{eff}$ = 2.22 eV is a
suitable choice to concurrently reproduce both the crystal structure and the electronic structure of the AFM FePS$_{3}$ monolayer.

\begin{figure}[thp]
    \centering
    \includegraphics[width=1.0\columnwidth]{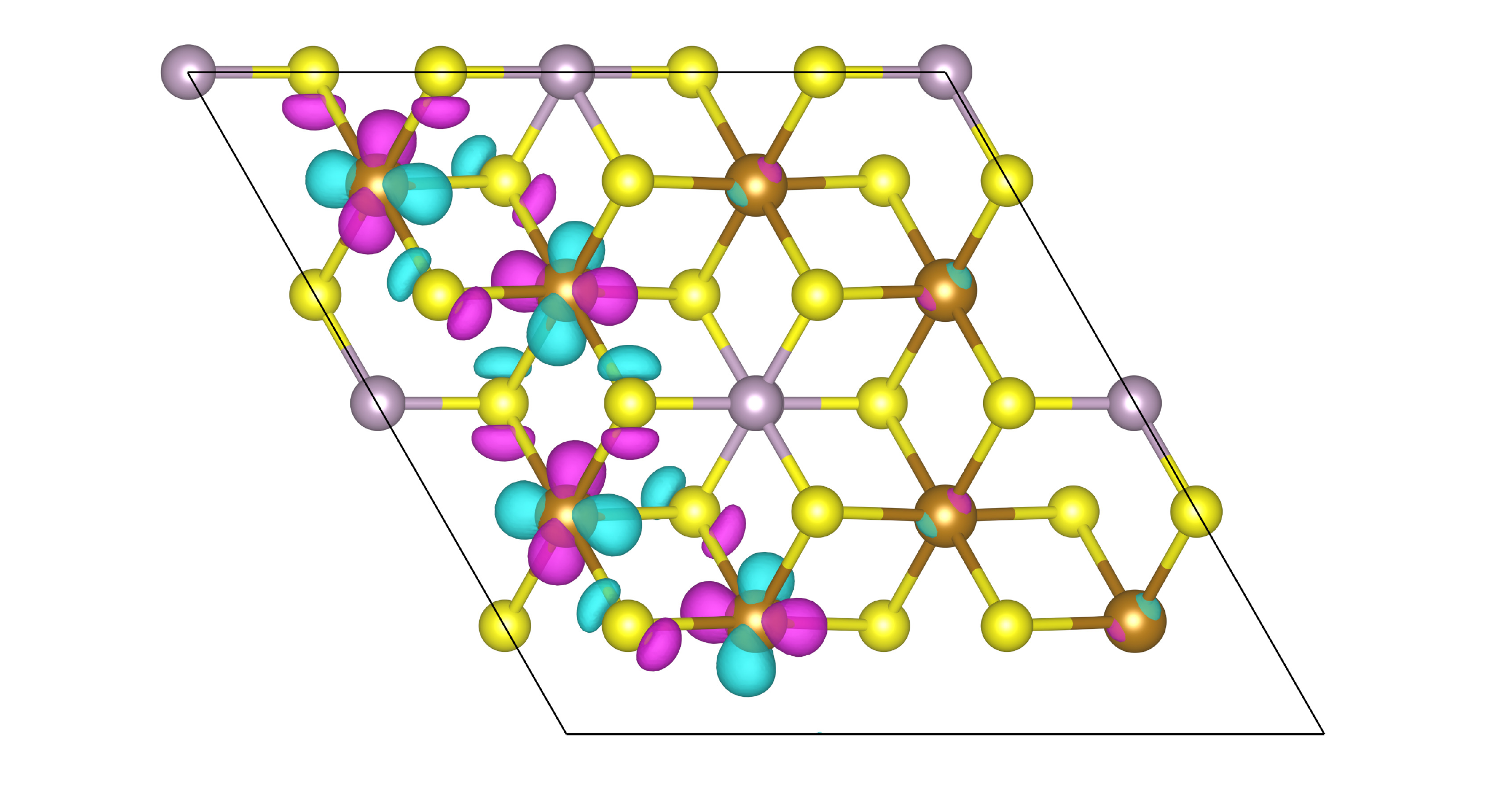}
   \caption{(Color online) Iso-contour plot of the wavefunction (shown only in the left part of the cell for clarity) or the highest valence band at the $\Gamma$ point. 
The overlap between the $d_{x^{2}-y^{2}}$ orbital of Fe ions  
and the $p$ orbitals of the S ions changes its sign. 
This is the reason why AFM and FM interactions alternate along the chain.
}
   \label{fig:charge-density}
\end{figure}

As indicated by the L{\"o}wdin charges in the $U_\text{eff}$ = 2.22~eV case, only a single orbital of the $3d$  shell, 
the in-plane $d_{x^{2}-y^{2}}$ orbital, is occupied. 
Thus, the electronic ground state of FePS$_3$ shows orbital ordering. 
To illustrate its role, 
a partial charge density plot of the highest occupied band at the $\Gamma$ point is displayed in Fig.~\ref{fig:charge-density}. 
It show that the spatial orientation of the in-plane Fe $3d$ orbital with respect to the crystal axes changes between the two Fe atoms in the structural unit cell.  In addition, sulfur $p$ orbitals oriented within the plane are involved in forming this electronic band. 
The alternating rotation of the Fe $3d$ orbital allows for a  bonding overlap (same sign of the orbital lobes) between the S atom and both Fe atoms for the long Fe -- Fe distance in the distorted hexagon. 
This finding explains why the long-bond zigzag chain is energetically preferred. 
%%%%%%%%%%%%%%%%%%%%%%%%%%%%%%%%%%%%%%%%%%%%%%%%%%%%%%%%%%%%%%%%%%%%%%%%%%%%%%%%%%%%%%%%%%%%%%%%%%%%%%%%%%%%%%%%%%%%%%%
%%%%%##################################################################################################################
Another aspect of orbital ordering is the unusually large orbital moment of Fe in this system. 
Using the Berry curvature approach~\cite{Malashevich_2010}, we find the orbital moment of each Fe to equal 0.77 and 0.8~$\mu_{B}$ for $U_\text{eff}$=2.22~eV and 4.89~eV, respectively. 
We believe that the reason for such a large orbital moment can be understood in analogy to the orbital moment of a free Fe atom which is determined by Hund's rule. 
In this system, Fe$^{2+}$ is in a $3d^{6}$ state. We thus expect 4~$\mu_{B}$ as spin moment, and DFT yields 3.46~$\mu_{B}$ for $U_\text{eff}$=2.22 eV. 
The difference is related to induced magnetic moments at the S and P ions. Moreover, for  $3d^{6}$ the quantum numbers $m_{l}$ do not add up to zero, which means the orbital moment is not quenched and points perpendicular to the plane. 
As a consequence, the effect of SOC and its related exchange interactions cannot be neglected, and there is a strong tendency for the spin to align with the orbital moment perpendicular to the plane. 

Because of the sensitivity to orbital ordering, we found out that great care must be taken when converging the electronic self-consistency cycle in the DFT$+U$ calculation. 
In particular, the choice of the starting point for the density matrix used to represent the electronic state of the single spin-down electron in the DFT$+U$ scheme has an influence on the converged result and must be checked carefully. % (see appendix).  
%%%%%%%%%%%%%%%%%%%%%%%%%%%%%%%%%%%%%%%%%%%%%%%%%%%%%%%%%%%%%%%%%%%%%%%%%%%%%%%%%%%%%%%%%%%%%%%%%%%%%%%%%%%%%%%%%%%%%%%%%%%5
\begin{table*} [!tbp]
  \centering
  \caption{Calculated Heisenberg  couplings $J_i$ (meV) up to the forth neighbors, biquadratic exchange interaction $B$(meV), Dzyaloshinskii-Moriya exchange interaction $D$(meV) and single ion anisotropy $\Delta$ (meV) for different $U_\text{eff}$(eV) parameters. Negative and
    positive value denotes anti-ferromagnetic and ferromagnetic
    exchange interaction, respectively.
    Note that $|S|=1$ has been used in the definition of the spin Hamiltonian. Using the obtained couplings, we perform MC simulations to find the Curie-Weiss ($\theta_\text{CW}$) and Neel ($T_\text{N}$) temperatures (K). We compare to experimental data from Ref.~\cite{lee2016ising, wang2016raman, JERNBERG1984178, Joy1992}.
    }
      \begin{tabular}{ccccccccccc}
    \hline
    $U_\text{eff}$ & $J_\text{1a}$ & $J_\text{1b}$ & $J_\text{2}$& $J_\text{3}$ & $J_\text{4}$&$\Delta$&$D$&$B$&$T_\text{N}$&$\theta_\text{CW}$\\
    \hline

    4.89 &  $-3.48$  & 3.63  & $-0.67$ & $-3.62$ &0.29&0.71& $-0.57$ & $-1.00$ &48.88&$-67.25$ \\
    3.89 &  $-3.87$  & 3.98  & $-1.14$ & $-4.01$ &0.20&0.73& $-0.39$ &$-1.37$ &51.42&$-73.12$ \\
    2.89 &  $-3.80$  & 4.13  & $-1.30$ & $-5.11$ &0.66&0.80& $-0.36$ & $-1.80$ &63.80&$-85.01$ \\
    2.22 &  $-3.26$  & 4.01  & $-1.24$ & $-5.71$ &1.50&0.89& $-0.34$ &$-2.10$ &70.00&$-101.57$ \\
    \hline
    exp. &       &       &       &       &        &       &       &      &104 -- 120&$-112$\\
     \hline
    \label{tab2}
  \end{tabular}
\end{table*}
%%%%%%%%%%%%%%%%%%%%%%%%%%%%%%%%%%%%%%%%%%%%%%%%%%%%%%%%%%%%%%%%%%%%%%%%%%%%%%%%%%%%%%%%%%%%%%%%%%%%%%%%%%%%%%%%%%%%%%%
\subsection{Magnetic exchange interactions} 
Next we map the information obtained from our DFT+U calculations onto a Heisenberg Hamiltonian
\begin{align}
H_{\rm {spin}} & = -\frac{1}{2}\sum_{i\neq j} J_{ij}(\vec{S_{i}}\cdot\vec{S_{j}})
\end{align}
Motivated by the orbital ordering leading to 'long' and 'short" distances between Fe neighbors, we allow for a differentiation of the first-neighbor $J$ parameters into $J_{1a}$ (= 'short') and $J_{1b}$ (= 'long') interactions. 
Since previous calculations had indicated a rather long range of the exchange interactions, despite the semiconducting character of FePS$_3$, we decided to include interactions up to the forth-nearest neighbors. 
Therefore, collinear magnetic configurations were calculated in a quite large $(2 \times 2)$ supercell using the QE code. 
The total energies for all possible 15 collinear structures were calculated to find the global minimum. 
The six configurations lowest in energy  shown in Fig.~\ref{fig:Hzn} were used to determine five exchange parameters. 
Note that, unlike in previous work \cite{Chittari,Olsen} the so-called 'stripy' magnetic pattern was not included in the fit because our calculations indicate that its energy is much higher that those of the other configurations.   

\subsection{Spin Hamiltonian and Monte Carlo simulation}
Now we proceed to discuss the features of the effective spin Hamiltonian, to find the ground state in a large simulation cell, and to calculate the finite-temperature properties of FePS$_{3}$ monolayer. 
To avoid any ambiguity due to the debatable value of the $U$ parameter, 
we investigate the dependence of the properties on this parameter systematically, varying it between the value determined from the band gap, $U=2.22$eV, and the value from DFPT, $U=4.89$ eV. 
Table~\ref{tab2} summarizes calculated $H$ terms for the different $U_\text{eff}$ parameters, as well as T$_\text{N}$ and Curie-Weiss temperature ($\theta_\text{CW}$), which  has been obtained by means of MC simulations using the Hamiltonian of  Eq.~\ref{H}.  

For all $U_\text{eff}$ parameters, exchange interaction for $J_{1a}$ and $J_{1b}$ is AFM and FM, respectively. This means that FM chains run along the long Fe -- Fe bond and couple to each other antiferromagnetically along the short bond. This is in agreement with the long-bond zigzag ground state~\cite{le1982magnetic}. 
Interestingly, some interactions are antiferromagnetic ($J_{1a}$ and $J_3$, negative sign), while others are ferromagnetic ($J_{1b}$ and $J_4$, positive sign).  
The value of $J_4$ is the smallest, thus ensuring that the magnetic exchange interaction falls off with distance, as one would expect in an insulating material. 
In absolute terms, $J_3$ has the largest value,  larger than $J_{1a}$. 
The dominance of $J_3$, which is responsible for the preferred antiparallel spin alignment between neighboring chains is getting even more prominent when a small value of $U$ (as dictated by the experimentally known electronic band gap) is employed. 
Remarkably, the small difference between $d_{1a}$ and $d_{1b}$ goes along with a different sign of the interactions $J_{1a}$ and $J_{1b}$, the latter being ferromagnetic.  
While this may seem surprising at first, we note that also in other 2D magnetic systems, such as Cr-trihalides, it has been reported that the exchange parameter may change its sign if the bond distances and bond angle change even by small amounts~\cite{Sandhukan-Delin-PRB2022}. 
In a wider context, this can be seen as a consequence of the Kanamori-Anderson-Goodenough rules \cite{Ka,Anderson,Go} that emphasize the role of the bond angle at the anion connecting two cations. The governing principle is the dependence of the exchange interaction on the orientation of the anion $p$ orbital relative to the bond axis between the magnetic cations. 
For more long-ranged interactions, 
the connectivity of the lattice sites (mediated via the $3p$ orbitals of P and S ions)  governs the super-exchange mechanism. According to the strong-coupling  perturbation theory, adding an extra intermediate ion (site-connection) increases the order of the perturbative expansion. 
Therefore, the super-exchange interaction does not reach zero with increasing distances as quickly as in other materials, e.g. in oxides. 

%%%%%%%%%%%%%%%%%%%%%%%%%%%%%%%%%%%%%%%%%%%%%%%%%%%%%%%%%%%%%%%%%%%%%%%%%%%%%%%%%%%%%%%%%%%%%%%%%%%%%%%%%%%%%%%%%%%%%%%
\begin{figure}[!htp]
    \centering
    \includegraphics[width=1.0\columnwidth]{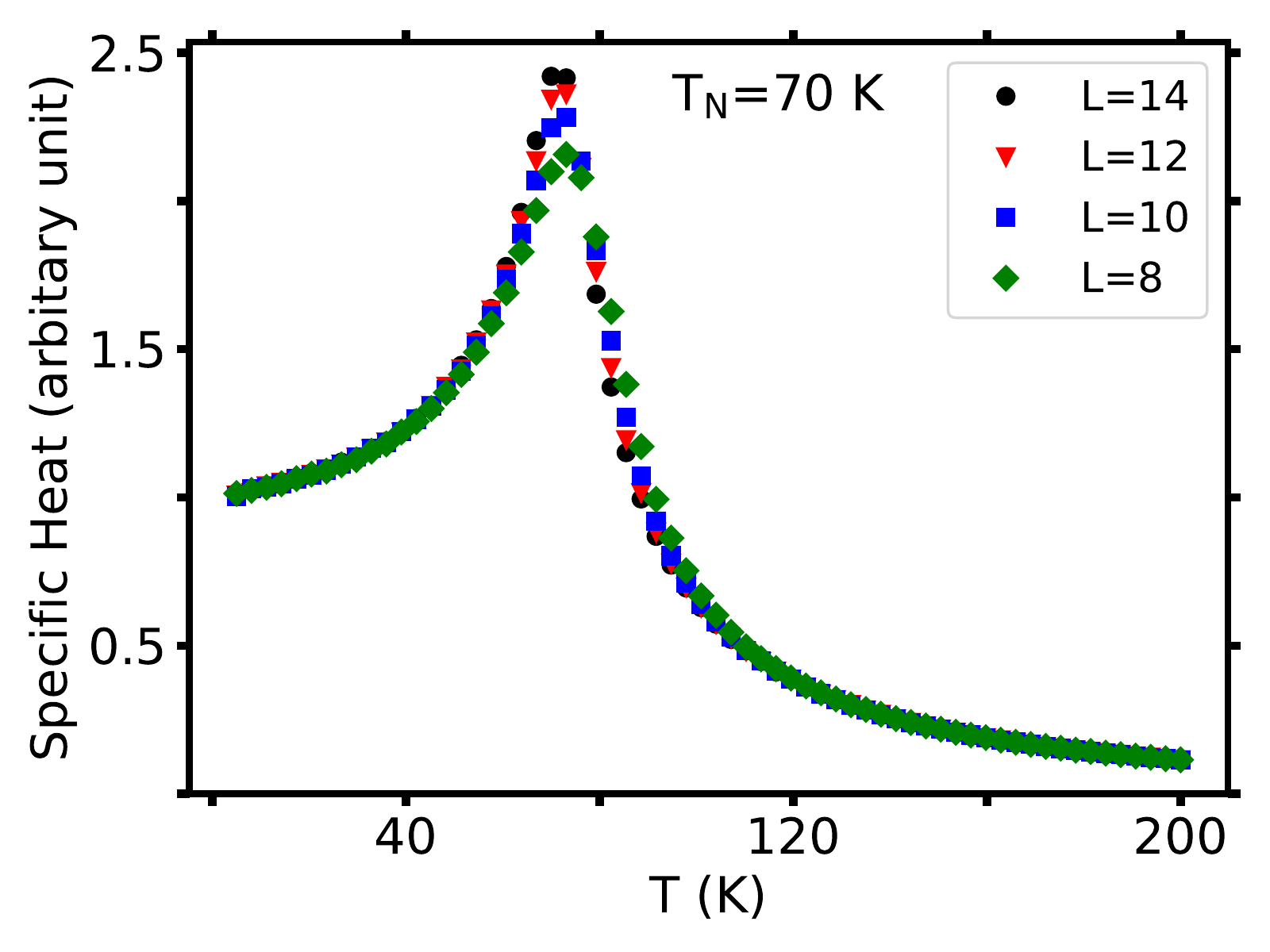}
   \caption{(Color online) Plot of the specific heat versus temperature obtained from Monte Carlo simulations of the spin Hamiltonian for lattices with the linear sizes $L=8$ to $L=14$ for $U_\text{eff}$=2.22 eV. 
   The divergence, seen as a peak of specific heat in the simulations, is used to read off the Neel temperature. 
   The observation that the peak height increases with increasing $L$ is indicative of a phase transition.}
   \label{fig:cv}
\end{figure}
%%%%%##################################################################################################################
As can be seen from Tab.~\ref{tab2}, the strengths of $\Delta$ and $D$ increase with decreasing $U_\text{eff}$ which satisfies Anderson's rule~\cite{Anderson}. 
Previous reports~\cite{Wildes_2012, Lancon2016, Olsen} didn't mention the effect of DMI in this system. They considered only the effect of the out-of-plane SIA which is responsible for the symmetry breaking required to have an ordered ground state in a 2D system. 
Here, our calculations show that the direction of the $D$ vector is along the $c$-axis due to the lack of inversion symmetry and helps SIA to break rotational symmetry. Our MC simulations show that without considering the effect of $\Delta$ and $D$, at finite temperature not all spins are oriented perpendicular to the plane.  
Besides increasing T$_\text{N}$, the effect of $\Delta$ and $D$ is to create perfect Ising-AFM order with spins pointing along the $c$-axis.
The negative value of $B$ favours the collinear coupling of spins. This feature also helps the system to settle in the Ising-AFM state. The presence of a biquadratic interaction in FePS$_{3}$ has been suggested by Wildes \textit{et al.}~\cite{wildes2020} when analyzing the experimental neutron scattering pattern and magnetoelastic effect. 

As a major result of the Monte Carlo simulations, we present the  
temperature dependence of the specific heat 
in Fig.~\ref{fig:cv} for $U_\text{eff}$=2.22 eV. The value of $T_\text{N}$ is estimated from the location of the peak. 
We observe that this peak becomes higher by increasing the simulation lattice size which is a confirmation of a phase transition taking place. 
We obtain the Curie-Weiss temperatures ($\theta_\text{CW}$) by a linear fitting of the inverse susceptibility in the temperature range between 150~K to 300~K. 
The negative values of $\theta_\text{CW}$ indicate that the ground state has AFM order. In addition, $\theta_\text{CW}$ increases monotonically with decreasing $U_\text{eff}$ which can be related to the enhancement of exchange interactions, in  good agreement with Anderson's theory of super-exchange~\cite{Anderson}. 
According to the definition\cite{Ramirez} of the frustration index $f$=$\frac{|\theta_\text{CW}|}{T_\text{N}}$, it is between 1 to 10 for all $U_\text{eff}$, indicating that this is a lightly frustrated system. 

While the obtained $\theta_{CW}$ is close to the experimental value, the obtained $T_\text{N}$ for $U_\text{eff}$=2.22 eV still falls short of the experimentally measured value. 
We note that other reports of first-principles based MC simulations for this material have also not been fully successful to obtain the exact Neel temperature~\cite{Olsen}.
A possible reason could be the neglect of 
spin-phonon coupling~\cite{webster2018distinct, Valenti, Xu2022} in our simulation. 
If FePS$_3$ undergoes a spin-Peierls transition, a gap in the magnetic excitation spectrum would result 
that is indeed observed in inelastic neutron scattering experiments\cite{Lancon2016}. 
Taking the gapped excitation spectrum into account, we expect a higher Neel temperature. 
However, simulations incorporating the quantum nature of joint phononic and magnetic excitations are beyond the scope of our current work.  

\section{Conclusion}
\label{sec:Conclusion}
In summary, using DFT calculations we determined a spin Hamiltonian for a single layer of FePS$_{3}$ 
with Heisenberg exchange up to the forth neighbor, the nearest-neighbor DM and biquadratic interactions and single-ion anisotropy. 
The three latter terms jointly result in the preferred alignment of Fe magnetic moments perpendicular to the crystallographic plane which has led to the characterization of FePS$_3$ as an Ising antiferromagnet.  
A series of calculations with different values $U_\text{eff}$ of the on-site Coulomb interactions in the Fe $3d$ shell has been carried out to ensure the robustness of the results. 
All calculations show that due to orbital ordering the honeycomb lattice formed by the Fe atoms is not ideal; two opposite sides of the Fe hexagons are longer than the others. 
We observe that calculated magnetic exchange couplings along the long bonds and short bonds are AFM and FM, respectively. 
As a consequence, the magnetic ground state must be described by a unit cell with (at least) four Fe atoms and consists of chains of parallel magnetic moments running across the long bonds. 
In an attempt to reproduce both the electronic band gap and the Neel temperature, we conclude that a value of $U_\text{eff}=2.22$eV is a reasonable choice, 
but there is room for improvement by including more subtle effects such as spin-lattice coupling.  
\section{acknowledgment}
We gratefully acknowledge the computing time granted by the Center for Computational Sciences and Simulation (CCSS) of the University of Duisburg-Essen and provided on the supercomputer magnitUDE (DFG Grant No. INST 20876/209-1 FUGG and INST 20876/243-1 FUGG) at the Zentrum f\"ur Informations- und Mediendienste (ZIM).
M. A. was supported by a fellowship from Universit\"at Duisburg-Essen 
and thanks his colleagues who provided insight and expertise that greatly assisted the research.
The guidelines provided by Dr. Hyun-Jung Kim and Dr. Gustav Bhilmayer  are gratefully acknowledged by M. A.

%

%\bibliographystyle{apsrev4-1}
%\bibliography{bib}

\begin{thebibliography}{51}%
\makeatletter
\providecommand \@ifxundefined [1]{%
 \@ifx{#1\undefined}
}%
\providecommand \@ifnum [1]{%
 \ifnum #1\expandafter \@firstoftwo
 \else \expandafter \@secondoftwo
 \fi
}%
\providecommand \@ifx [1]{%
 \ifx #1\expandafter \@firstoftwo
 \else \expandafter \@secondoftwo
 \fi
}%
\providecommand \natexlab [1]{#1}%
\providecommand \enquote  [1]{``#1''}%
\providecommand \bibnamefont  [1]{#1}%
\providecommand \bibfnamefont [1]{#1}%
\providecommand \citenamefont [1]{#1}%
\providecommand \href@noop [0]{\@secondoftwo}%
\providecommand \href [0]{\begingroup \@sanitize@url \@href}%
\providecommand \@href[1]{\@@startlink{#1}\@@href}%
\providecommand \@@href[1]{\endgroup#1\@@endlink}%
\providecommand \@sanitize@url [0]{\catcode `\\12\catcode `\$12\catcode
  `\&12\catcode `\#12\catcode `\^12\catcode `\_12\catcode `\%12\relax}%
\providecommand \@@startlink[1]{}%
\providecommand \@@endlink[0]{}%
\providecommand \url  [0]{\begingroup\@sanitize@url \@url }%
\providecommand \@url [1]{\endgroup\@href {#1}{\urlprefix }}%
\providecommand \urlprefix  [0]{URL }%
\providecommand \Eprint [0]{\href }%
\providecommand \doibase [0]{http://dx.doi.org/}%
\providecommand \selectlanguage [0]{\@gobble}%
\providecommand \bibinfo  [0]{\@secondoftwo}%
\providecommand \bibfield  [0]{\@secondoftwo}%
\providecommand \translation [1]{[#1]}%
\providecommand \BibitemOpen [0]{}%
\providecommand \bibitemStop [0]{}%
\providecommand \bibitemNoStop [0]{.\EOS\space}%
\providecommand \EOS [0]{\spacefactor3000\relax}%
\providecommand \BibitemShut  [1]{\csname bibitem#1\endcsname}%
\let\auto@bib@innerbib\@empty
%</preamble>
\bibitem [{\citenamefont {Huang}\ \emph {et~al.}(2017)\citenamefont {Huang},
  \citenamefont {Clark}, \citenamefont {Navarro-Moratalla}, \citenamefont
  {Klein}, \citenamefont {Cheng}, \citenamefont {Seyler}, \citenamefont
  {Zhong}, \citenamefont {Schmidgall}, \citenamefont {McGuire}, \citenamefont
  {Cobden}, \citenamefont {Yao}, \citenamefont {Xiao}, \citenamefont
  {Jarillo-Herrero},\ and\ \citenamefont {Xu}}]{Huang2017}%
  \BibitemOpen
  \bibfield  {author} {\bibinfo {author} {\bibfnamefont {B.}~\bibnamefont
  {Huang}}, \bibinfo {author} {\bibfnamefont {G.}~\bibnamefont {Clark}},
  \bibinfo {author} {\bibfnamefont {E.}~\bibnamefont {Navarro-Moratalla}},
  \bibinfo {author} {\bibfnamefont {D.~R.}\ \bibnamefont {Klein}}, \bibinfo
  {author} {\bibfnamefont {R.}~\bibnamefont {Cheng}}, \bibinfo {author}
  {\bibfnamefont {K.~L.}\ \bibnamefont {Seyler}}, \bibinfo {author}
  {\bibfnamefont {D.}~\bibnamefont {Zhong}}, \bibinfo {author} {\bibfnamefont
  {E.}~\bibnamefont {Schmidgall}}, \bibinfo {author} {\bibfnamefont {M.~A.}\
  \bibnamefont {McGuire}}, \bibinfo {author} {\bibfnamefont {D.~H.}\
  \bibnamefont {Cobden}}, \bibinfo {author} {\bibfnamefont {W.}~\bibnamefont
  {Yao}}, \bibinfo {author} {\bibfnamefont {D.}~\bibnamefont {Xiao}}, \bibinfo
  {author} {\bibfnamefont {P.}~\bibnamefont {Jarillo-Herrero}}, \ and\ \bibinfo
  {author} {\bibfnamefont {X.}~\bibnamefont {Xu}},\ }\href {\doibase
  10.1038/nature22391} {\bibfield  {journal} {\bibinfo  {journal} {Nature}\
  }\textbf {\bibinfo {volume} {546}},\ \bibinfo {pages} {270} (\bibinfo {year}
  {2017})}\BibitemShut {NoStop}%
\bibitem [{\citenamefont {Gong}\ \emph {et~al.}(2017)\citenamefont {Gong},
  \citenamefont {Li}, \citenamefont {Li}, \citenamefont {Ji}, \citenamefont
  {Stern}, \citenamefont {Xia}, \citenamefont {Cao}, \citenamefont {Bao},
  \citenamefont {Wang}, \citenamefont {Wang}, \citenamefont {Qiu},
  \citenamefont {Cava}, \citenamefont {Louie}, \citenamefont {Xia},\ and\
  \citenamefont {Zhang}}]{Gong2017}%
  \BibitemOpen
  \bibfield  {author} {\bibinfo {author} {\bibfnamefont {C.}~\bibnamefont
  {Gong}}, \bibinfo {author} {\bibfnamefont {L.}~\bibnamefont {Li}}, \bibinfo
  {author} {\bibfnamefont {Z.}~\bibnamefont {Li}}, \bibinfo {author}
  {\bibfnamefont {H.}~\bibnamefont {Ji}}, \bibinfo {author} {\bibfnamefont
  {A.}~\bibnamefont {Stern}}, \bibinfo {author} {\bibfnamefont
  {Y.}~\bibnamefont {Xia}}, \bibinfo {author} {\bibfnamefont {T.}~\bibnamefont
  {Cao}}, \bibinfo {author} {\bibfnamefont {W.}~\bibnamefont {Bao}}, \bibinfo
  {author} {\bibfnamefont {C.}~\bibnamefont {Wang}}, \bibinfo {author}
  {\bibfnamefont {Y.}~\bibnamefont {Wang}}, \bibinfo {author} {\bibfnamefont
  {Z.~Q.}\ \bibnamefont {Qiu}}, \bibinfo {author} {\bibfnamefont {R.~J.}\
  \bibnamefont {Cava}}, \bibinfo {author} {\bibfnamefont {S.~G.}\ \bibnamefont
  {Louie}}, \bibinfo {author} {\bibfnamefont {J.}~\bibnamefont {Xia}}, \ and\
  \bibinfo {author} {\bibfnamefont {X.}~\bibnamefont {Zhang}},\ }\href
  {\doibase 10.1038/nature22060} {\bibfield  {journal} {\bibinfo  {journal}
  {Nature}\ }\textbf {\bibinfo {volume} {546}},\ \bibinfo {pages} {265}
  (\bibinfo {year} {2017})}\BibitemShut {NoStop}%
\bibitem [{\citenamefont {Chhowalla}\ \emph {et~al.}(2013)\citenamefont
  {Chhowalla}, \citenamefont {Shin}, \citenamefont {Eda}, \citenamefont {Li},
  \citenamefont {Loh},\ and\ \citenamefont {Zhang}}]{Chhowalla2013}%
  \BibitemOpen
  \bibfield  {author} {\bibinfo {author} {\bibfnamefont {M.}~\bibnamefont
  {Chhowalla}}, \bibinfo {author} {\bibfnamefont {H.~S.}\ \bibnamefont {Shin}},
  \bibinfo {author} {\bibfnamefont {G.}~\bibnamefont {Eda}}, \bibinfo {author}
  {\bibfnamefont {L.-J.}\ \bibnamefont {Li}}, \bibinfo {author} {\bibfnamefont
  {K.~P.}\ \bibnamefont {Loh}}, \ and\ \bibinfo {author} {\bibfnamefont
  {H.}~\bibnamefont {Zhang}},\ }\href {\doibase 10.1038/nchem.1589} {\bibfield
  {journal} {\bibinfo  {journal} {Nature Chemistry}\ }\textbf {\bibinfo
  {volume} {5}},\ \bibinfo {pages} {263} (\bibinfo {year} {2013})}\BibitemShut
  {NoStop}%
\bibitem [{\citenamefont {Zhang}\ \emph {et~al.}(2015)\citenamefont {Zhang},
  \citenamefont {Qu}, \citenamefont {Zhu},\ and\ \citenamefont
  {Lam}}]{Zhang2015}%
  \BibitemOpen
  \bibfield  {author} {\bibinfo {author} {\bibfnamefont {W.-B.}\ \bibnamefont
  {Zhang}}, \bibinfo {author} {\bibfnamefont {Q.}~\bibnamefont {Qu}}, \bibinfo
  {author} {\bibfnamefont {P.}~\bibnamefont {Zhu}}, \ and\ \bibinfo {author}
  {\bibfnamefont {C.-H.}\ \bibnamefont {Lam}},\ }\href {\doibase
  10.1039/C5TC02840J} {\bibfield  {journal} {\bibinfo  {journal} {J. Mater.
  Chem. C}\ }\textbf {\bibinfo {volume} {3}},\ \bibinfo {pages} {12457}
  (\bibinfo {year} {2015})}\BibitemShut {NoStop}%
\bibitem [{\citenamefont {Ahn}(2020)}]{Ahn2020}%
  \BibitemOpen
  \bibfield  {author} {\bibinfo {author} {\bibfnamefont {E.~C.}\ \bibnamefont
  {Ahn}},\ }\href {\doibase 10.1038/s41699-020-0152-0} {\bibfield  {journal}
  {\bibinfo  {journal} {npj 2D Materials and Applications}\ }\textbf {\bibinfo
  {volume} {4}},\ \bibinfo {pages} {17} (\bibinfo {year} {2020})}\BibitemShut
  {NoStop}%
\bibitem [{\citenamefont {Lee}\ \emph {et~al.}(2016)\citenamefont {Lee},
  \citenamefont {Lee}, \citenamefont {Ryoo}, \citenamefont {Kang},
  \citenamefont {Kim}, \citenamefont {Kim}, \citenamefont {Park}, \citenamefont
  {Park},\ and\ \citenamefont {Cheong}}]{lee2016ising}%
  \BibitemOpen
  \bibfield  {author} {\bibinfo {author} {\bibfnamefont {J.-U.}\ \bibnamefont
  {Lee}}, \bibinfo {author} {\bibfnamefont {S.}~\bibnamefont {Lee}}, \bibinfo
  {author} {\bibfnamefont {J.~H.}\ \bibnamefont {Ryoo}}, \bibinfo {author}
  {\bibfnamefont {S.}~\bibnamefont {Kang}}, \bibinfo {author} {\bibfnamefont
  {T.~Y.}\ \bibnamefont {Kim}}, \bibinfo {author} {\bibfnamefont
  {P.}~\bibnamefont {Kim}}, \bibinfo {author} {\bibfnamefont {C.-H.}\
  \bibnamefont {Park}}, \bibinfo {author} {\bibfnamefont {J.-G.}\ \bibnamefont
  {Park}}, \ and\ \bibinfo {author} {\bibfnamefont {H.}~\bibnamefont
  {Cheong}},\ }\href {\doibase 10.1021/acs.nanolett.6b03052} {\bibfield
  {journal} {\bibinfo  {journal} {Nano Letters}\ }\textbf {\bibinfo {volume}
  {16}},\ \bibinfo {pages} {7433} (\bibinfo {year} {2016})}\BibitemShut
  {NoStop}%
\bibitem [{\citenamefont {Wang}\ \emph {et~al.}(2016)\citenamefont {Wang},
  \citenamefont {Du}, \citenamefont {Liu}, \citenamefont {Hu}, \citenamefont
  {Zhang}, \citenamefont {Zhang}, \citenamefont {Owen}, \citenamefont {Lu},
  \citenamefont {Gan}, \citenamefont {Sengupta}, \citenamefont {Kloc},\ and\
  \citenamefont {Xiong}}]{wang2016raman}%
  \BibitemOpen
  \bibfield  {author} {\bibinfo {author} {\bibfnamefont {X.}~\bibnamefont
  {Wang}}, \bibinfo {author} {\bibfnamefont {K.}~\bibnamefont {Du}}, \bibinfo
  {author} {\bibfnamefont {Y.~Y.~F.}\ \bibnamefont {Liu}}, \bibinfo {author}
  {\bibfnamefont {P.}~\bibnamefont {Hu}}, \bibinfo {author} {\bibfnamefont
  {J.}~\bibnamefont {Zhang}}, \bibinfo {author} {\bibfnamefont
  {Q.}~\bibnamefont {Zhang}}, \bibinfo {author} {\bibfnamefont {M.~H.~S.}\
  \bibnamefont {Owen}}, \bibinfo {author} {\bibfnamefont {X.}~\bibnamefont
  {Lu}}, \bibinfo {author} {\bibfnamefont {C.~K.}\ \bibnamefont {Gan}},
  \bibinfo {author} {\bibfnamefont {P.}~\bibnamefont {Sengupta}}, \bibinfo
  {author} {\bibfnamefont {C.}~\bibnamefont {Kloc}}, \ and\ \bibinfo {author}
  {\bibfnamefont {Q.}~\bibnamefont {Xiong}},\ }\href {\doibase
  10.1088/2053-1583/3/3/031009} {\bibfield  {journal} {\bibinfo  {journal} {2D
  Materials}\ }\textbf {\bibinfo {volume} {3}},\ \bibinfo {pages} {031009}
  (\bibinfo {year} {2016})}\BibitemShut {NoStop}%
\bibitem [{\citenamefont {Ramos}\ \emph {et~al.}(2021)\citenamefont {Ramos},
  \citenamefont {Carrascoso}, \citenamefont {Frisenda}, \citenamefont {Gant},
  \citenamefont {Ma{\~{n}}as-Valero}, \citenamefont {Esteras}, \citenamefont
  {Baldov{\'i}}, \citenamefont {Coronado}, \citenamefont {Castellanos-Gomez},\
  and\ \citenamefont {Calvo}}]{Ramos2021}%
  \BibitemOpen
  \bibfield  {author} {\bibinfo {author} {\bibfnamefont {M.}~\bibnamefont
  {Ramos}}, \bibinfo {author} {\bibfnamefont {F.}~\bibnamefont {Carrascoso}},
  \bibinfo {author} {\bibfnamefont {R.}~\bibnamefont {Frisenda}}, \bibinfo
  {author} {\bibfnamefont {P.}~\bibnamefont {Gant}}, \bibinfo {author}
  {\bibfnamefont {S.}~\bibnamefont {Ma{\~{n}}as-Valero}}, \bibinfo {author}
  {\bibfnamefont {D.~L.}\ \bibnamefont {Esteras}}, \bibinfo {author}
  {\bibfnamefont {J.~J.}\ \bibnamefont {Baldov{\'i}}}, \bibinfo {author}
  {\bibfnamefont {E.}~\bibnamefont {Coronado}}, \bibinfo {author}
  {\bibfnamefont {A.}~\bibnamefont {Castellanos-Gomez}}, \ and\ \bibinfo
  {author} {\bibfnamefont {M.~R.}\ \bibnamefont {Calvo}},\ }\href {\doibase
  10.1038/s41699-021-00199-z} {\bibfield  {journal} {\bibinfo  {journal} {npj
  2D Materials and Applications}\ }\textbf {\bibinfo {volume} {5}},\ \bibinfo
  {pages} {19} (\bibinfo {year} {2021})}\BibitemShut {NoStop}%
\bibitem [{\citenamefont {Haines}\ \emph {et~al.}(2018)\citenamefont {Haines},
  \citenamefont {Coak}, \citenamefont {Wildes}, \citenamefont {Lampronti},
  \citenamefont {Liu}, \citenamefont {Nahai-Williamson}, \citenamefont
  {Hamidov}, \citenamefont {Daisenberger},\ and\ \citenamefont
  {Saxena}}]{Haines2018}%
  \BibitemOpen
  \bibfield  {author} {\bibinfo {author} {\bibfnamefont {C.~R.~S.}\
  \bibnamefont {Haines}}, \bibinfo {author} {\bibfnamefont {M.~J.}\
  \bibnamefont {Coak}}, \bibinfo {author} {\bibfnamefont {A.~R.}\ \bibnamefont
  {Wildes}}, \bibinfo {author} {\bibfnamefont {G.~I.}\ \bibnamefont
  {Lampronti}}, \bibinfo {author} {\bibfnamefont {C.}~\bibnamefont {Liu}},
  \bibinfo {author} {\bibfnamefont {P.}~\bibnamefont {Nahai-Williamson}},
  \bibinfo {author} {\bibfnamefont {H.}~\bibnamefont {Hamidov}}, \bibinfo
  {author} {\bibfnamefont {D.}~\bibnamefont {Daisenberger}}, \ and\ \bibinfo
  {author} {\bibfnamefont {S.~S.}\ \bibnamefont {Saxena}},\ }\href {\doibase
  10.1103/PhysRevLett.121.266801} {\bibfield  {journal} {\bibinfo  {journal}
  {Phys. Rev. Lett.}\ }\textbf {\bibinfo {volume} {121}},\ \bibinfo {pages}
  {266801} (\bibinfo {year} {2018})}\BibitemShut {NoStop}%
\bibitem [{\citenamefont {Brec}\ \emph {et~al.}(1979)\citenamefont {Brec},
  \citenamefont {Schleich}, \citenamefont {Ouvrard}, \citenamefont {Louisy},\
  and\ \citenamefont {Rouxel}}]{Brec1979}%
  \BibitemOpen
  \bibfield  {author} {\bibinfo {author} {\bibfnamefont {R.}~\bibnamefont
  {Brec}}, \bibinfo {author} {\bibfnamefont {D.~M.}\ \bibnamefont {Schleich}},
  \bibinfo {author} {\bibfnamefont {G.}~\bibnamefont {Ouvrard}}, \bibinfo
  {author} {\bibfnamefont {A.}~\bibnamefont {Louisy}}, \ and\ \bibinfo {author}
  {\bibfnamefont {J.}~\bibnamefont {Rouxel}},\ }\href {\doibase
  10.1021/ic50197a018} {\bibfield  {journal} {\bibinfo  {journal} {Inorganic
  Chemistry}\ }\textbf {\bibinfo {volume} {18}},\ \bibinfo {pages} {1814}
  (\bibinfo {year} {1979})}\BibitemShut {NoStop}%
\bibitem [{\citenamefont {Foot}\ \emph {et~al.}(1980)\citenamefont {Foot},
  \citenamefont {Suradi},\ and\ \citenamefont {Lee}}]{FOOT1980189}%
  \BibitemOpen
  \bibfield  {author} {\bibinfo {author} {\bibfnamefont {P.}~\bibnamefont
  {Foot}}, \bibinfo {author} {\bibfnamefont {J.}~\bibnamefont {Suradi}}, \ and\
  \bibinfo {author} {\bibfnamefont {P.}~\bibnamefont {Lee}},\ }\href {\doibase
  10.1016/0025-5408(80)90118-X} {\bibfield  {journal} {\bibinfo  {journal}
  {Materials Research Bulletin}\ }\textbf {\bibinfo {volume} {15}},\ \bibinfo
  {pages} {189} (\bibinfo {year} {1980})}\BibitemShut {NoStop}%
\bibitem [{\citenamefont {Gao}\ \emph {et~al.}(2018)\citenamefont {Gao},
  \citenamefont {Lei}, \citenamefont {Kang}, \citenamefont {Fei}, \citenamefont
  {Mak}, \citenamefont {Yuan}, \citenamefont {Zhang}, \citenamefont {Li},
  \citenamefont {Bao}, \citenamefont {Zeng}, \citenamefont {Wang},
  \citenamefont {Gu},\ and\ \citenamefont {Zhang}}]{Gao_2018}%
  \BibitemOpen
  \bibfield  {author} {\bibinfo {author} {\bibfnamefont {Y.}~\bibnamefont
  {Gao}}, \bibinfo {author} {\bibfnamefont {S.}~\bibnamefont {Lei}}, \bibinfo
  {author} {\bibfnamefont {T.}~\bibnamefont {Kang}}, \bibinfo {author}
  {\bibfnamefont {L.}~\bibnamefont {Fei}}, \bibinfo {author} {\bibfnamefont
  {C.-L.}\ \bibnamefont {Mak}}, \bibinfo {author} {\bibfnamefont
  {J.}~\bibnamefont {Yuan}}, \bibinfo {author} {\bibfnamefont {M.}~\bibnamefont
  {Zhang}}, \bibinfo {author} {\bibfnamefont {S.}~\bibnamefont {Li}}, \bibinfo
  {author} {\bibfnamefont {Q.}~\bibnamefont {Bao}}, \bibinfo {author}
  {\bibfnamefont {Z.}~\bibnamefont {Zeng}}, \bibinfo {author} {\bibfnamefont
  {Z.}~\bibnamefont {Wang}}, \bibinfo {author} {\bibfnamefont {H.}~\bibnamefont
  {Gu}}, \ and\ \bibinfo {author} {\bibfnamefont {K.}~\bibnamefont {Zhang}},\
  }\href {\doibase 10.1088/1361-6528/aab9d2} {\bibfield  {journal} {\bibinfo
  {journal} {Nanotechnology}\ }\textbf {\bibinfo {volume} {29}},\ \bibinfo
  {pages} {244001} (\bibinfo {year} {2018})}\BibitemShut {NoStop}%
\bibitem [{\citenamefont {Xu}\ \emph {et~al.}(2020)\citenamefont {Xu},
  \citenamefont {Guo}, \citenamefont {Tu}, \citenamefont {Li}, \citenamefont
  {Chen}, \citenamefont {Chen}, \citenamefont {Tian}, \citenamefont {Chen},
  \citenamefont {Shi}, \citenamefont {Li}, \citenamefont {Su},\ and\
  \citenamefont {Fan}}]{xu2020controllable}%
  \BibitemOpen
  \bibfield  {author} {\bibinfo {author} {\bibfnamefont {D.}~\bibnamefont
  {Xu}}, \bibinfo {author} {\bibfnamefont {Z.}~\bibnamefont {Guo}}, \bibinfo
  {author} {\bibfnamefont {Y.}~\bibnamefont {Tu}}, \bibinfo {author}
  {\bibfnamefont {X.}~\bibnamefont {Li}}, \bibinfo {author} {\bibfnamefont
  {Y.}~\bibnamefont {Chen}}, \bibinfo {author} {\bibfnamefont {Z.}~\bibnamefont
  {Chen}}, \bibinfo {author} {\bibfnamefont {B.}~\bibnamefont {Tian}}, \bibinfo
  {author} {\bibfnamefont {S.}~\bibnamefont {Chen}}, \bibinfo {author}
  {\bibfnamefont {Y.}~\bibnamefont {Shi}}, \bibinfo {author} {\bibfnamefont
  {Y.}~\bibnamefont {Li}}, \bibinfo {author} {\bibfnamefont {C.}~\bibnamefont
  {Su}}, \ and\ \bibinfo {author} {\bibfnamefont {D.}~\bibnamefont {Fan}},\
  }\href {\doibase 10.1515/nanoph-2020-0336} {\bibfield  {journal} {\bibinfo
  {journal} {Nanophotonics}\ }\textbf {\bibinfo {volume} {9}},\ \bibinfo
  {pages} {4555} (\bibinfo {year} {2020})}\BibitemShut {NoStop}%
\bibitem [{\citenamefont {Lan\ifmmode~\mbox{\c{c}}\else \c{c}\fi{}on}\ \emph
  {et~al.}(2016)\citenamefont {Lan\ifmmode~\mbox{\c{c}}\else \c{c}\fi{}on},
  \citenamefont {Walker}, \citenamefont {Ressouche}, \citenamefont {Ouladdiaf},
  \citenamefont {Rule}, \citenamefont {McIntyre}, \citenamefont {Hicks},
  \citenamefont {R\o{}nnow},\ and\ \citenamefont {Wildes}}]{Lancon2016}%
  \BibitemOpen
  \bibfield  {author} {\bibinfo {author} {\bibfnamefont {D.}~\bibnamefont
  {Lan\ifmmode~\mbox{\c{c}}\else \c{c}\fi{}on}}, \bibinfo {author}
  {\bibfnamefont {H.~C.}\ \bibnamefont {Walker}}, \bibinfo {author}
  {\bibfnamefont {E.}~\bibnamefont {Ressouche}}, \bibinfo {author}
  {\bibfnamefont {B.}~\bibnamefont {Ouladdiaf}}, \bibinfo {author}
  {\bibfnamefont {K.~C.}\ \bibnamefont {Rule}}, \bibinfo {author}
  {\bibfnamefont {G.~J.}\ \bibnamefont {McIntyre}}, \bibinfo {author}
  {\bibfnamefont {T.~J.}\ \bibnamefont {Hicks}}, \bibinfo {author}
  {\bibfnamefont {H.~M.}\ \bibnamefont {R\o{}nnow}}, \ and\ \bibinfo {author}
  {\bibfnamefont {A.~R.}\ \bibnamefont {Wildes}},\ }\href {\doibase
  10.1103/PhysRevB.94.214407} {\bibfield  {journal} {\bibinfo  {journal} {Phys.
  Rev. B}\ }\textbf {\bibinfo {volume} {94}},\ \bibinfo {pages} {214407}
  (\bibinfo {year} {2016})}\BibitemShut {NoStop}%
\bibitem [{\citenamefont {Cheng}\ \emph {et~al.}(2021)\citenamefont {Cheng},
  \citenamefont {Lee}, \citenamefont {Iyer}, \citenamefont {Chica},
  \citenamefont {Qian}, \citenamefont {Shehzad}, \citenamefont {dos Reis},
  \citenamefont {Kanatzidis},\ and\ \citenamefont {Dravid}}]{Cheng2021}%
  \BibitemOpen
  \bibfield  {author} {\bibinfo {author} {\bibfnamefont {M.}~\bibnamefont
  {Cheng}}, \bibinfo {author} {\bibfnamefont {Y.-S.}\ \bibnamefont {Lee}},
  \bibinfo {author} {\bibfnamefont {A.~K.}\ \bibnamefont {Iyer}}, \bibinfo
  {author} {\bibfnamefont {D.~G.}\ \bibnamefont {Chica}}, \bibinfo {author}
  {\bibfnamefont {E.~K.}\ \bibnamefont {Qian}}, \bibinfo {author}
  {\bibfnamefont {M.~A.}\ \bibnamefont {Shehzad}}, \bibinfo {author}
  {\bibfnamefont {R.}~\bibnamefont {dos Reis}}, \bibinfo {author}
  {\bibfnamefont {M.~G.}\ \bibnamefont {Kanatzidis}}, \ and\ \bibinfo {author}
  {\bibfnamefont {V.~P.}\ \bibnamefont {Dravid}},\ }\href {\doibase
  10.1021/acs.inorgchem.1c02635} {\bibfield  {journal} {\bibinfo  {journal}
  {Inorganic Chemistry}\ }\textbf {\bibinfo {volume} {60}},\ \bibinfo {pages}
  {17268} (\bibinfo {year} {2021})}\BibitemShut {NoStop}%
\bibitem [{\citenamefont {Momma}\ and\ \citenamefont {Izumi}(2011)}]{vesta}%
  \BibitemOpen
  \bibfield  {author} {\bibinfo {author} {\bibfnamefont {K.}~\bibnamefont
  {Momma}}\ and\ \bibinfo {author} {\bibfnamefont {F.}~\bibnamefont {Izumi}},\
  }\href {\doibase 10.1107/S0021889811038970} {\bibfield  {journal} {\bibinfo
  {journal} {Journal of Applied Crystallography}\ }\textbf {\bibinfo {volume}
  {44}},\ \bibinfo {pages} {1272} (\bibinfo {year} {2011})}\BibitemShut
  {NoStop}%
\bibitem [{\citenamefont {Klingen}\ \emph {et~al.}(1973)\citenamefont
  {Klingen}, \citenamefont {Eulenberger},\ and\ \citenamefont
  {Hahn}}]{Klingen1973}%
  \BibitemOpen
  \bibfield  {author} {\bibinfo {author} {\bibfnamefont {W.}~\bibnamefont
  {Klingen}}, \bibinfo {author} {\bibfnamefont {G.}~\bibnamefont
  {Eulenberger}}, \ and\ \bibinfo {author} {\bibfnamefont {H.}~\bibnamefont
  {Hahn}},\ }\href {\doibase 10.1002/zaac.19734010113} {\bibfield  {journal}
  {\bibinfo  {journal} {Zeitschrift fur anorganische und allgemeine Chemie}\
  }\textbf {\bibinfo {volume} {401}},\ \bibinfo {pages} {97} (\bibinfo {year}
  {1973})}\BibitemShut {NoStop}%
\bibitem [{\citenamefont {Mermin}\ and\ \citenamefont
  {Wagner}(1966)}]{Mermin1966}%
  \BibitemOpen
  \bibfield  {author} {\bibinfo {author} {\bibfnamefont {N.~D.}\ \bibnamefont
  {Mermin}}\ and\ \bibinfo {author} {\bibfnamefont {H.}~\bibnamefont
  {Wagner}},\ }\href {\doibase 10.1103/PhysRevLett.17.1307} {\bibfield
  {journal} {\bibinfo  {journal} {Phys. Rev. Lett.}\ }\textbf {\bibinfo
  {volume} {17}},\ \bibinfo {pages} {1307} (\bibinfo {year}
  {1966})}\BibitemShut {NoStop}%
\bibitem [{\citenamefont {Hohenberg}(1967)}]{Hohenberg1967}%
  \BibitemOpen
  \bibfield  {author} {\bibinfo {author} {\bibfnamefont {P.~C.}\ \bibnamefont
  {Hohenberg}},\ }\href {\doibase 10.1103/PhysRev.158.383} {\bibfield
  {journal} {\bibinfo  {journal} {Phys. Rev.}\ }\textbf {\bibinfo {volume}
  {158}},\ \bibinfo {pages} {383} (\bibinfo {year} {1967})}\BibitemShut
  {NoStop}%
\bibitem [{\citenamefont {Olsen}(2021)}]{Olsen}%
  \BibitemOpen
  \bibfield  {author} {\bibinfo {author} {\bibfnamefont {T.}~\bibnamefont
  {Olsen}},\ }\href {\doibase 10.1088/1361-6463/ac000e} {\bibfield  {journal}
  {\bibinfo  {journal} {Journal of Physics D: Applied Physics}\ }\textbf
  {\bibinfo {volume} {54}},\ \bibinfo {pages} {314001} (\bibinfo {year}
  {2021})}\BibitemShut {NoStop}%
\bibitem [{\citenamefont {Chittari}\ \emph {et~al.}(2016)\citenamefont
  {Chittari}, \citenamefont {Park}, \citenamefont {Lee}, \citenamefont {Han},
  \citenamefont {MacDonald}, \citenamefont {Hwang},\ and\ \citenamefont
  {Jung}}]{Chittari}%
  \BibitemOpen
  \bibfield  {author} {\bibinfo {author} {\bibfnamefont {B.~L.}\ \bibnamefont
  {Chittari}}, \bibinfo {author} {\bibfnamefont {Y.}~\bibnamefont {Park}},
  \bibinfo {author} {\bibfnamefont {D.}~\bibnamefont {Lee}}, \bibinfo {author}
  {\bibfnamefont {M.}~\bibnamefont {Han}}, \bibinfo {author} {\bibfnamefont
  {A.~H.}\ \bibnamefont {MacDonald}}, \bibinfo {author} {\bibfnamefont
  {E.}~\bibnamefont {Hwang}}, \ and\ \bibinfo {author} {\bibfnamefont
  {J.}~\bibnamefont {Jung}},\ }\href {\doibase 10.1103/PhysRevB.94.184428}
  {\bibfield  {journal} {\bibinfo  {journal} {Phys. Rev. B}\ }\textbf {\bibinfo
  {volume} {94}},\ \bibinfo {pages} {184428} (\bibinfo {year}
  {2016})}\BibitemShut {NoStop}%
\bibitem [{\citenamefont {Kurosawa}\ \emph {et~al.}(1983)\citenamefont
  {Kurosawa}, \citenamefont {Saito},\ and\ \citenamefont
  {Yamaguchi}}]{Kurosawa}%
  \BibitemOpen
  \bibfield  {author} {\bibinfo {author} {\bibfnamefont {K.}~\bibnamefont
  {Kurosawa}}, \bibinfo {author} {\bibfnamefont {S.}~\bibnamefont {Saito}}, \
  and\ \bibinfo {author} {\bibfnamefont {Y.}~\bibnamefont {Yamaguchi}},\ }\href
  {\doibase 10.1143/JPSJ.52.3919} {\bibfield  {journal} {\bibinfo  {journal}
  {Journal of the Physical Society of Japan}\ }\textbf {\bibinfo {volume}
  {52}},\ \bibinfo {pages} {3919} (\bibinfo {year} {1983})}\BibitemShut
  {NoStop}%
\bibitem [{\citenamefont {{Le Flem}}\ \emph {et~al.}(1982)\citenamefont {{Le
  Flem}}, \citenamefont {Brec}, \citenamefont {Ouvard}, \citenamefont
  {Louisy},\ and\ \citenamefont {Segransan}}]{le1982magnetic}%
  \BibitemOpen
  \bibfield  {author} {\bibinfo {author} {\bibfnamefont {G.}~\bibnamefont {{Le
  Flem}}}, \bibinfo {author} {\bibfnamefont {R.}~\bibnamefont {Brec}}, \bibinfo
  {author} {\bibfnamefont {G.}~\bibnamefont {Ouvard}}, \bibinfo {author}
  {\bibfnamefont {A.}~\bibnamefont {Louisy}}, \ and\ \bibinfo {author}
  {\bibfnamefont {P.}~\bibnamefont {Segransan}},\ }\href {\doibase
  10.1016/0022-3697(82)90156-1} {\bibfield  {journal} {\bibinfo  {journal}
  {Journal of Physics and Chemistry of Solids}\ }\textbf {\bibinfo {volume}
  {43}},\ \bibinfo {pages} {455} (\bibinfo {year} {1982})}\BibitemShut
  {NoStop}%
\bibitem [{\citenamefont {min Zhang}\ \emph {et~al.}(2021)\citenamefont {min
  Zhang}, \citenamefont {zhuang Nie}, \citenamefont {guang Wang}, \citenamefont
  {lin Xia},\ and\ \citenamefont {hua Guo}}]{ZHANG2021167687}%
  \BibitemOpen
  \bibfield  {author} {\bibinfo {author} {\bibfnamefont {J.}~\bibnamefont {min
  Zhang}}, \bibinfo {author} {\bibfnamefont {Y.}~\bibnamefont {zhuang Nie}},
  \bibinfo {author} {\bibfnamefont {X.}~\bibnamefont {guang Wang}}, \bibinfo
  {author} {\bibfnamefont {Q.}~\bibnamefont {lin Xia}}, \ and\ \bibinfo
  {author} {\bibfnamefont {G.}~\bibnamefont {hua Guo}},\ }\href {\doibase
  https://doi.org/10.1016/j.jmmm.2020.167687} {\bibfield  {journal} {\bibinfo
  {journal} {Journal of Magnetism and Magnetic Materials}\ }\textbf {\bibinfo
  {volume} {525}},\ \bibinfo {pages} {167687} (\bibinfo {year}
  {2021})}\BibitemShut {NoStop}%
\bibitem [{\citenamefont {Wildes}\ \emph {et~al.}(2012)\citenamefont {Wildes},
  \citenamefont {Rule}, \citenamefont {Bewley}, \citenamefont {Enderle},\ and\
  \citenamefont {Hicks}}]{Wildes_2012}%
  \BibitemOpen
  \bibfield  {author} {\bibinfo {author} {\bibfnamefont {A.~R.}\ \bibnamefont
  {Wildes}}, \bibinfo {author} {\bibfnamefont {K.~C.}\ \bibnamefont {Rule}},
  \bibinfo {author} {\bibfnamefont {R.~I.}\ \bibnamefont {Bewley}}, \bibinfo
  {author} {\bibfnamefont {M.}~\bibnamefont {Enderle}}, \ and\ \bibinfo
  {author} {\bibfnamefont {T.~J.}\ \bibnamefont {Hicks}},\ }\href {\doibase
  10.1088/0953-8984/24/41/416004} {\bibfield  {journal} {\bibinfo  {journal}
  {Journal of Physics: Condensed Matter}\ }\textbf {\bibinfo {volume} {24}},\
  \bibinfo {pages} {416004} (\bibinfo {year} {2012})}\BibitemShut {NoStop}%
\bibitem [{\citenamefont {Rule}\ \emph {et~al.}(2007)\citenamefont {Rule},
  \citenamefont {McIntyre}, \citenamefont {Kennedy},\ and\ \citenamefont
  {Hicks}}]{Rule2007}%
  \BibitemOpen
  \bibfield  {author} {\bibinfo {author} {\bibfnamefont {K.~C.}\ \bibnamefont
  {Rule}}, \bibinfo {author} {\bibfnamefont {G.~J.}\ \bibnamefont {McIntyre}},
  \bibinfo {author} {\bibfnamefont {S.~J.}\ \bibnamefont {Kennedy}}, \ and\
  \bibinfo {author} {\bibfnamefont {T.~J.}\ \bibnamefont {Hicks}},\ }\href
  {\doibase 10.1103/PhysRevB.76.134402} {\bibfield  {journal} {\bibinfo
  {journal} {Phys. Rev. B}\ }\textbf {\bibinfo {volume} {76}},\ \bibinfo
  {pages} {134402} (\bibinfo {year} {2007})}\BibitemShut {NoStop}%
\bibitem [{\citenamefont {Wildes}\ \emph {et~al.}(2020)\citenamefont {Wildes},
  \citenamefont {Zhitomirsky}, \citenamefont {Ziman}, \citenamefont {Lancon},\
  and\ \citenamefont {Walker}}]{wildes2020}%
  \BibitemOpen
  \bibfield  {author} {\bibinfo {author} {\bibfnamefont {A.~R.}\ \bibnamefont
  {Wildes}}, \bibinfo {author} {\bibfnamefont {M.~E.}\ \bibnamefont
  {Zhitomirsky}}, \bibinfo {author} {\bibfnamefont {T.}~\bibnamefont {Ziman}},
  \bibinfo {author} {\bibfnamefont {D.}~\bibnamefont {Lancon}}, \ and\ \bibinfo
  {author} {\bibfnamefont {H.~C.}\ \bibnamefont {Walker}},\ }\href {\doibase
  10.1063/5.0009114} {\bibfield  {journal} {\bibinfo  {journal} {Journal of
  Applied Physics}\ }\textbf {\bibinfo {volume} {127}},\ \bibinfo {pages}
  {223903} (\bibinfo {year} {2020})}\BibitemShut {NoStop}%
\bibitem [{\citenamefont {Murayama}\ \emph {et~al.}(2016)\citenamefont
  {Murayama}, \citenamefont {Okabe}, \citenamefont {Urushihara}, \citenamefont
  {Asaka}, \citenamefont {Fukuda}, \citenamefont {Isobe}, \citenamefont
  {Yamamoto},\ and\ \citenamefont {Matsushita}}]{Murayama2016}%
  \BibitemOpen
  \bibfield  {author} {\bibinfo {author} {\bibfnamefont {C.}~\bibnamefont
  {Murayama}}, \bibinfo {author} {\bibfnamefont {M.}~\bibnamefont {Okabe}},
  \bibinfo {author} {\bibfnamefont {D.}~\bibnamefont {Urushihara}}, \bibinfo
  {author} {\bibfnamefont {T.}~\bibnamefont {Asaka}}, \bibinfo {author}
  {\bibfnamefont {K.}~\bibnamefont {Fukuda}}, \bibinfo {author} {\bibfnamefont
  {M.}~\bibnamefont {Isobe}}, \bibinfo {author} {\bibfnamefont
  {K.}~\bibnamefont {Yamamoto}}, \ and\ \bibinfo {author} {\bibfnamefont
  {Y.}~\bibnamefont {Matsushita}},\ }\href {\doibase 10.1063/1.4961712}
  {\bibfield  {journal} {\bibinfo  {journal} {Journal of Applied Physics}\
  }\textbf {\bibinfo {volume} {120}},\ \bibinfo {pages} {142114} (\bibinfo
  {year} {2016})}\BibitemShut {NoStop}%
\bibitem [{\citenamefont {Ouvrard}\ \emph {et~al.}(1985)\citenamefont
  {Ouvrard}, \citenamefont {Brec},\ and\ \citenamefont
  {Rouxel}}]{OUVRARD19851181}%
  \BibitemOpen
  \bibfield  {author} {\bibinfo {author} {\bibfnamefont {G.}~\bibnamefont
  {Ouvrard}}, \bibinfo {author} {\bibfnamefont {R.}~\bibnamefont {Brec}}, \
  and\ \bibinfo {author} {\bibfnamefont {J.}~\bibnamefont {Rouxel}},\ }\href
  {\doibase https://doi.org/10.1016/0025-5408(85)90092-3} {\bibfield  {journal}
  {\bibinfo  {journal} {Materials Research Bulletin}\ }\textbf {\bibinfo
  {volume} {20}},\ \bibinfo {pages} {1181} (\bibinfo {year}
  {1985})}\BibitemShut {NoStop}%
\bibitem [{\citenamefont {Li}\ \emph {et~al.}(2021)\citenamefont {Li},
  \citenamefont {Yu}, \citenamefont {Lou}, \citenamefont {Feng}, \citenamefont
  {Whangbo},\ and\ \citenamefont {Xiang}}]{li2021spin}%
  \BibitemOpen
  \bibfield  {author} {\bibinfo {author} {\bibfnamefont {X.}~\bibnamefont
  {Li}}, \bibinfo {author} {\bibfnamefont {H.}~\bibnamefont {Yu}}, \bibinfo
  {author} {\bibfnamefont {F.}~\bibnamefont {Lou}}, \bibinfo {author}
  {\bibfnamefont {J.}~\bibnamefont {Feng}}, \bibinfo {author} {\bibfnamefont
  {M.-H.}\ \bibnamefont {Whangbo}}, \ and\ \bibinfo {author} {\bibfnamefont
  {H.}~\bibnamefont {Xiang}},\ }\href {\doibase 10.3390/molecules26040803}
  {\bibfield  {journal} {\bibinfo  {journal} {Molecules}\ }\textbf {\bibinfo
  {volume} {26}} (\bibinfo {year} {2021}),\
  10.3390/molecules26040803}\BibitemShut {NoStop}%
\bibitem [{\citenamefont {Sadeghi}\ \emph {et~al.}(2015)\citenamefont
  {Sadeghi}, \citenamefont {Alaei}, \citenamefont {Shahbazi},\ and\
  \citenamefont {Gingras}}]{Sadeghi2015}%
  \BibitemOpen
  \bibfield  {author} {\bibinfo {author} {\bibfnamefont {A.}~\bibnamefont
  {Sadeghi}}, \bibinfo {author} {\bibfnamefont {M.}~\bibnamefont {Alaei}},
  \bibinfo {author} {\bibfnamefont {F.}~\bibnamefont {Shahbazi}}, \ and\
  \bibinfo {author} {\bibfnamefont {M.~J.~P.}\ \bibnamefont {Gingras}},\ }\href
  {\doibase 10.1103/PhysRevB.91.140407} {\bibfield  {journal} {\bibinfo
  {journal} {Phys. Rev. B}\ }\textbf {\bibinfo {volume} {91}},\ \bibinfo
  {pages} {140407} (\bibinfo {year} {2015})}\BibitemShut {NoStop}%
\bibitem [{\citenamefont {Giannozzi}\ \emph {et~al.}(2009)\citenamefont
  {Giannozzi}, \citenamefont {Baroni}, \citenamefont {Bonini}, \citenamefont
  {Calandra}, \citenamefont {Car}, \citenamefont {Cavazzoni}, \citenamefont
  {Ceresoli}, \citenamefont {Chiarotti}, \citenamefont {Cococcioni},
  \citenamefont {Dabo}, \citenamefont {Corso}, \citenamefont {de~Gironcoli},
  \citenamefont {Fabris}, \citenamefont {Fratesi}, \citenamefont {Gebauer},
  \citenamefont {Gerstmann}, \citenamefont {Gougoussis}, \citenamefont
  {Kokalj}, \citenamefont {Lazzeri}, \citenamefont {Martin-Samos},
  \citenamefont {Marzari}, \citenamefont {Mauri}, \citenamefont {Mazzarello},
  \citenamefont {Paolini}, \citenamefont {Pasquarello}, \citenamefont
  {Paulatto}, \citenamefont {Sbraccia}, \citenamefont {Scandolo}, \citenamefont
  {Sclauzero}, \citenamefont {Seitsonen}, \citenamefont {Smogunov},
  \citenamefont {Umari},\ and\ \citenamefont {Wentzcovitch}}]{Giannozzi_2009}%
  \BibitemOpen
  \bibfield  {author} {\bibinfo {author} {\bibfnamefont {P.}~\bibnamefont
  {Giannozzi}}, \bibinfo {author} {\bibfnamefont {S.}~\bibnamefont {Baroni}},
  \bibinfo {author} {\bibfnamefont {N.}~\bibnamefont {Bonini}}, \bibinfo
  {author} {\bibfnamefont {M.}~\bibnamefont {Calandra}}, \bibinfo {author}
  {\bibfnamefont {R.}~\bibnamefont {Car}}, \bibinfo {author} {\bibfnamefont
  {C.}~\bibnamefont {Cavazzoni}}, \bibinfo {author} {\bibfnamefont
  {D.}~\bibnamefont {Ceresoli}}, \bibinfo {author} {\bibfnamefont {G.~L.}\
  \bibnamefont {Chiarotti}}, \bibinfo {author} {\bibfnamefont {M.}~\bibnamefont
  {Cococcioni}}, \bibinfo {author} {\bibfnamefont {I.}~\bibnamefont {Dabo}},
  \bibinfo {author} {\bibfnamefont {A.~D.}\ \bibnamefont {Corso}}, \bibinfo
  {author} {\bibfnamefont {S.}~\bibnamefont {de~Gironcoli}}, \bibinfo {author}
  {\bibfnamefont {S.}~\bibnamefont {Fabris}}, \bibinfo {author} {\bibfnamefont
  {G.}~\bibnamefont {Fratesi}}, \bibinfo {author} {\bibfnamefont
  {R.}~\bibnamefont {Gebauer}}, \bibinfo {author} {\bibfnamefont
  {U.}~\bibnamefont {Gerstmann}}, \bibinfo {author} {\bibfnamefont
  {C.}~\bibnamefont {Gougoussis}}, \bibinfo {author} {\bibfnamefont
  {A.}~\bibnamefont {Kokalj}}, \bibinfo {author} {\bibfnamefont
  {M.}~\bibnamefont {Lazzeri}}, \bibinfo {author} {\bibfnamefont
  {L.}~\bibnamefont {Martin-Samos}}, \bibinfo {author} {\bibfnamefont
  {N.}~\bibnamefont {Marzari}}, \bibinfo {author} {\bibfnamefont
  {F.}~\bibnamefont {Mauri}}, \bibinfo {author} {\bibfnamefont
  {R.}~\bibnamefont {Mazzarello}}, \bibinfo {author} {\bibfnamefont
  {S.}~\bibnamefont {Paolini}}, \bibinfo {author} {\bibfnamefont
  {A.}~\bibnamefont {Pasquarello}}, \bibinfo {author} {\bibfnamefont
  {L.}~\bibnamefont {Paulatto}}, \bibinfo {author} {\bibfnamefont
  {C.}~\bibnamefont {Sbraccia}}, \bibinfo {author} {\bibfnamefont
  {S.}~\bibnamefont {Scandolo}}, \bibinfo {author} {\bibfnamefont
  {G.}~\bibnamefont {Sclauzero}}, \bibinfo {author} {\bibfnamefont {A.~P.}\
  \bibnamefont {Seitsonen}}, \bibinfo {author} {\bibfnamefont {A.}~\bibnamefont
  {Smogunov}}, \bibinfo {author} {\bibfnamefont {P.}~\bibnamefont {Umari}}, \
  and\ \bibinfo {author} {\bibfnamefont {R.~M.}\ \bibnamefont {Wentzcovitch}},\
  }\href {\doibase 10.1088/0953-8984/21/39/395502} {\bibfield  {journal}
  {\bibinfo  {journal} {Journal of Physics: Condensed Matter}\ }\textbf
  {\bibinfo {volume} {21}},\ \bibinfo {pages} {395502} (\bibinfo {year}
  {2009})}\BibitemShut {NoStop}%
\bibitem [{\citenamefont {Perdew}\ \emph {et~al.}(1996)\citenamefont {Perdew},
  \citenamefont {Burke},\ and\ \citenamefont {Ernzerhof}}]{Perdew1996}%
  \BibitemOpen
  \bibfield  {author} {\bibinfo {author} {\bibfnamefont {J.~P.}\ \bibnamefont
  {Perdew}}, \bibinfo {author} {\bibfnamefont {K.}~\bibnamefont {Burke}}, \
  and\ \bibinfo {author} {\bibfnamefont {M.}~\bibnamefont {Ernzerhof}},\ }\href
  {\doibase 10.1103/PhysRevLett.77.3865} {\bibfield  {journal} {\bibinfo
  {journal} {Phys. Rev. Lett.}\ }\textbf {\bibinfo {volume} {77}},\ \bibinfo
  {pages} {3865} (\bibinfo {year} {1996})}\BibitemShut {NoStop}%
\bibitem [{\citenamefont {Garrity}\ \emph {et~al.}(2014)\citenamefont
  {Garrity}, \citenamefont {Bennett}, \citenamefont {Rabe},\ and\ \citenamefont
  {Vanderbilt}}]{Vanderbilt2014}%
  \BibitemOpen
  \bibfield  {author} {\bibinfo {author} {\bibfnamefont {K.~F.}\ \bibnamefont
  {Garrity}}, \bibinfo {author} {\bibfnamefont {J.~W.}\ \bibnamefont
  {Bennett}}, \bibinfo {author} {\bibfnamefont {K.~M.}\ \bibnamefont {Rabe}}, \
  and\ \bibinfo {author} {\bibfnamefont {D.}~\bibnamefont {Vanderbilt}},\
  }\href {\doibase 10.1016/j.commatsci.2013.08.053} {\bibfield  {journal}
  {\bibinfo  {journal} {Computational Materials Science}\ }\textbf {\bibinfo
  {volume} {81}},\ \bibinfo {pages} {446} (\bibinfo {year} {2014})}\BibitemShut
  {NoStop}%
\bibitem [{\citenamefont {FLEURgroup}()}]{fleur}%
  \BibitemOpen
  \bibfield  {author} {\bibinfo {author} {\bibnamefont {FLEURgroup}},\ }\href
  {http://www.flapw.de/} {\enquote {\bibinfo {title} {http://www.flapw.de/},}\
  }\BibitemShut {NoStop}%
\bibitem [{\citenamefont {Moriya}(1960)}]{Moriya1960}%
  \BibitemOpen
  \bibfield  {author} {\bibinfo {author} {\bibfnamefont {T.}~\bibnamefont
  {Moriya}},\ }\href {\doibase 10.1103/PhysRev.120.91} {\bibfield  {journal}
  {\bibinfo  {journal} {Phys. Rev.}\ }\textbf {\bibinfo {volume} {120}},\
  \bibinfo {pages} {91} (\bibinfo {year} {1960})}\BibitemShut {NoStop}%
\bibitem [{\citenamefont {Mila}\ and\ \citenamefont {Zhang}(2000)}]{Mila2000}%
  \BibitemOpen
  \bibfield  {author} {\bibinfo {author} {\bibfnamefont {F.}~\bibnamefont
  {Mila}}\ and\ \bibinfo {author} {\bibfnamefont {F.-C.}\ \bibnamefont
  {Zhang}},\ }\href {\doibase 10.1007/s100510070242} {\bibfield  {journal}
  {\bibinfo  {journal} {The European Physical Journal B - Condensed Matter and
  Complex Systems}\ }\textbf {\bibinfo {volume} {16}},\ \bibinfo {pages} {7}
  (\bibinfo {year} {2000})}\BibitemShut {NoStop}%
\bibitem [{\citenamefont {Tanaka}\ \emph {et~al.}(2018)\citenamefont {Tanaka},
  \citenamefont {Yokoyama},\ and\ \citenamefont {Hotta}}]{Tanaka2018}%
  \BibitemOpen
  \bibfield  {author} {\bibinfo {author} {\bibfnamefont {K.}~\bibnamefont
  {Tanaka}}, \bibinfo {author} {\bibfnamefont {Y.}~\bibnamefont {Yokoyama}}, \
  and\ \bibinfo {author} {\bibfnamefont {C.}~\bibnamefont {Hotta}},\ }\href
  {\doibase 10.7566/JPSJ.87.023702} {\bibfield  {journal} {\bibinfo  {journal}
  {Journal of the Physical Society of Japan}\ }\textbf {\bibinfo {volume}
  {87}},\ \bibinfo {pages} {023702} (\bibinfo {year} {2018})}\BibitemShut
  {NoStop}%
\bibitem [{\citenamefont {Hukushima}\ and\ \citenamefont
  {Nemoto}(1996)}]{Hukushima1996}%
  \BibitemOpen
  \bibfield  {author} {\bibinfo {author} {\bibfnamefont {K.}~\bibnamefont
  {Hukushima}}\ and\ \bibinfo {author} {\bibfnamefont {K.}~\bibnamefont
  {Nemoto}},\ }\href {\doibase 10.1143/JPSJ.65.1604} {\bibfield  {journal}
  {\bibinfo  {journal} {Journal of the Physical Society of Japan}\ }\textbf
  {\bibinfo {volume} {65}},\ \bibinfo {pages} {1604} (\bibinfo {year}
  {1996})}\BibitemShut {NoStop}%
\bibitem [{\citenamefont {Timrov}\ \emph {et~al.}(2018)\citenamefont {Timrov},
  \citenamefont {Marzari},\ and\ \citenamefont {Cococcioni}}]{Timrov2018}%
  \BibitemOpen
  \bibfield  {author} {\bibinfo {author} {\bibfnamefont {I.}~\bibnamefont
  {Timrov}}, \bibinfo {author} {\bibfnamefont {N.}~\bibnamefont {Marzari}}, \
  and\ \bibinfo {author} {\bibfnamefont {M.}~\bibnamefont {Cococcioni}},\
  }\href {\doibase 10.1103/PhysRevB.98.085127} {\bibfield  {journal} {\bibinfo
  {journal} {Phys. Rev. B}\ }\textbf {\bibinfo {volume} {98}},\ \bibinfo
  {pages} {085127} (\bibinfo {year} {2018})}\BibitemShut {NoStop}%
\bibitem [{\citenamefont {Malashevich}\ \emph {et~al.}(2010)\citenamefont
  {Malashevich}, \citenamefont {Souza}, \citenamefont {Coh},\ and\
  \citenamefont {Vanderbilt}}]{Malashevich_2010}%
  \BibitemOpen
  \bibfield  {author} {\bibinfo {author} {\bibfnamefont {A.}~\bibnamefont
  {Malashevich}}, \bibinfo {author} {\bibfnamefont {I.}~\bibnamefont {Souza}},
  \bibinfo {author} {\bibfnamefont {S.}~\bibnamefont {Coh}}, \ and\ \bibinfo
  {author} {\bibfnamefont {D.}~\bibnamefont {Vanderbilt}},\ }\href {\doibase
  10.1088/1367-2630/12/5/053032} {\bibfield  {journal} {\bibinfo  {journal}
  {New Journal of Physics}\ }\textbf {\bibinfo {volume} {12}},\ \bibinfo
  {pages} {053032} (\bibinfo {year} {2010})}\BibitemShut {NoStop}%
\bibitem [{\citenamefont {Jernberg}\ \emph {et~al.}(1984)\citenamefont
  {Jernberg}, \citenamefont {Bjarman},\ and\ \citenamefont
  {Wappling}}]{JERNBERG1984178}%
  \BibitemOpen
  \bibfield  {author} {\bibinfo {author} {\bibfnamefont {P.}~\bibnamefont
  {Jernberg}}, \bibinfo {author} {\bibfnamefont {S.}~\bibnamefont {Bjarman}}, \
  and\ \bibinfo {author} {\bibfnamefont {R.}~\bibnamefont {Wappling}},\ }\href
  {\doibase https://doi.org/10.1016/0304-8853(84)90355-X} {\bibfield  {journal}
  {\bibinfo  {journal} {Journal of Magnetism and Magnetic Materials}\ }\textbf
  {\bibinfo {volume} {46}},\ \bibinfo {pages} {178} (\bibinfo {year}
  {1984})}\BibitemShut {NoStop}%
\bibitem [{\citenamefont {Joy}\ and\ \citenamefont
  {Vasudevan}(1992)}]{Joy1992}%
  \BibitemOpen
  \bibfield  {author} {\bibinfo {author} {\bibfnamefont {P.~A.}\ \bibnamefont
  {Joy}}\ and\ \bibinfo {author} {\bibfnamefont {S.}~\bibnamefont
  {Vasudevan}},\ }\href {\doibase 10.1103/PhysRevB.46.5425} {\bibfield
  {journal} {\bibinfo  {journal} {Phys. Rev. B}\ }\textbf {\bibinfo {volume}
  {46}},\ \bibinfo {pages} {5425} (\bibinfo {year} {1992})}\BibitemShut
  {NoStop}%
\bibitem [{\citenamefont {Sadhukhan}\ \emph {et~al.}(2022)\citenamefont
  {Sadhukhan}, \citenamefont {Bergman}, \citenamefont {Kvashnin}, \citenamefont
  {Hellsvik},\ and\ \citenamefont {Delin}}]{Sandhukan-Delin-PRB2022}%
  \BibitemOpen
  \bibfield  {author} {\bibinfo {author} {\bibfnamefont {B.}~\bibnamefont
  {Sadhukhan}}, \bibinfo {author} {\bibfnamefont {A.}~\bibnamefont {Bergman}},
  \bibinfo {author} {\bibfnamefont {Y.~O.}\ \bibnamefont {Kvashnin}}, \bibinfo
  {author} {\bibfnamefont {J.}~\bibnamefont {Hellsvik}}, \ and\ \bibinfo
  {author} {\bibfnamefont {A.}~\bibnamefont {Delin}},\ }\href {\doibase
  10.1103/PhysRevB.105.104418} {\bibfield  {journal} {\bibinfo  {journal}
  {Phys. Rev. B}\ }\textbf {\bibinfo {volume} {105}},\ \bibinfo {pages}
  {104418} (\bibinfo {year} {2022})}\BibitemShut {NoStop}%
\bibitem [{\citenamefont {Kanamori}(1959)}]{Ka}%
  \BibitemOpen
  \bibfield  {author} {\bibinfo {author} {\bibfnamefont {J.}~\bibnamefont
  {Kanamori}},\ }\href {\doibase https://doi.org/10.1016/0022-3697(59)90061-7}
  {\bibfield  {journal} {\bibinfo  {journal} {Journal of Physics and Chemistry
  of Solids}\ }\textbf {\bibinfo {volume} {10}},\ \bibinfo {pages} {87}
  (\bibinfo {year} {1959})}\BibitemShut {NoStop}%
\bibitem [{\citenamefont {Anderson}(1950)}]{Anderson}%
  \BibitemOpen
  \bibfield  {author} {\bibinfo {author} {\bibfnamefont {P.~W.}\ \bibnamefont
  {Anderson}},\ }\href {\doibase 10.1103/PhysRev.79.350} {\bibfield  {journal}
  {\bibinfo  {journal} {Phys. Rev.}\ }\textbf {\bibinfo {volume} {79}},\
  \bibinfo {pages} {350} (\bibinfo {year} {1950})}\BibitemShut {NoStop}%
\bibitem [{\citenamefont {Goodenough}(1955)}]{Go}%
  \BibitemOpen
  \bibfield  {author} {\bibinfo {author} {\bibfnamefont {J.~B.}\ \bibnamefont
  {Goodenough}},\ }\href {\doibase 10.1103/PhysRev.100.564} {\bibfield
  {journal} {\bibinfo  {journal} {Phys. Rev.}\ }\textbf {\bibinfo {volume}
  {100}},\ \bibinfo {pages} {564} (\bibinfo {year} {1955})}\BibitemShut
  {NoStop}%
\bibitem [{\citenamefont {Ramirez}(1994)}]{Ramirez}%
  \BibitemOpen
  \bibfield  {author} {\bibinfo {author} {\bibfnamefont {A.~P.}\ \bibnamefont
  {Ramirez}},\ }\href {\doibase 10.1146/annurev.ms.24.080194.002321} {\bibfield
   {journal} {\bibinfo  {journal} {Annual Review of Materials Science}\
  }\textbf {\bibinfo {volume} {24}},\ \bibinfo {pages} {453} (\bibinfo {year}
  {1994})}\BibitemShut {NoStop}%
\bibitem [{\citenamefont {Webster}\ \emph {et~al.}(2018)\citenamefont
  {Webster}, \citenamefont {Liang},\ and\ \citenamefont
  {Yan}}]{webster2018distinct}%
  \BibitemOpen
  \bibfield  {author} {\bibinfo {author} {\bibfnamefont {L.}~\bibnamefont
  {Webster}}, \bibinfo {author} {\bibfnamefont {L.}~\bibnamefont {Liang}}, \
  and\ \bibinfo {author} {\bibfnamefont {J.-A.}\ \bibnamefont {Yan}},\
  }\href@noop {} {\bibfield  {journal} {\bibinfo  {journal} {Physical Chemistry
  Chemical Physics}\ }\textbf {\bibinfo {volume} {20}},\ \bibinfo {pages}
  {23546} (\bibinfo {year} {2018})}\BibitemShut {NoStop}%
\bibitem [{\citenamefont {Ferrari}\ \emph {et~al.}(2021)\citenamefont
  {Ferrari}, \citenamefont {Valent\'{\i}},\ and\ \citenamefont
  {Becca}}]{Valenti}%
  \BibitemOpen
  \bibfield  {author} {\bibinfo {author} {\bibfnamefont {F.}~\bibnamefont
  {Ferrari}}, \bibinfo {author} {\bibfnamefont {R.}~\bibnamefont
  {Valent\'{\i}}}, \ and\ \bibinfo {author} {\bibfnamefont {F.}~\bibnamefont
  {Becca}},\ }\href {\doibase 10.1103/PhysRevB.104.035126} {\bibfield
  {journal} {\bibinfo  {journal} {Phys. Rev. B}\ }\textbf {\bibinfo {volume}
  {104}},\ \bibinfo {pages} {035126} (\bibinfo {year} {2021})}\BibitemShut
  {NoStop}%
\bibitem [{\citenamefont {Xu}\ \emph {et~al.}(2022)\citenamefont {Xu},
  \citenamefont {Wang}, \citenamefont {Chang}, \citenamefont {Chen},
  \citenamefont {Guan},\ and\ \citenamefont {Tao}}]{Xu2022}%
  \BibitemOpen
  \bibfield  {author} {\bibinfo {author} {\bibfnamefont {X.}~\bibnamefont
  {Xu}}, \bibinfo {author} {\bibfnamefont {X.}~\bibnamefont {Wang}}, \bibinfo
  {author} {\bibfnamefont {P.}~\bibnamefont {Chang}}, \bibinfo {author}
  {\bibfnamefont {X.}~\bibnamefont {Chen}}, \bibinfo {author} {\bibfnamefont
  {L.}~\bibnamefont {Guan}}, \ and\ \bibinfo {author} {\bibfnamefont
  {J.}~\bibnamefont {Tao}},\ }\href {\doibase 10.1021/acs.jpcc.2c02742}
  {\bibfield  {journal} {\bibinfo  {journal} {The Journal of Physical Chemistry
  C}\ }\textbf {\bibinfo {volume} {126}},\ \bibinfo {pages} {10574} (\bibinfo
  {year} {2022})}\BibitemShut {NoStop}%
\end{thebibliography}
\end{document}